\documentclass[prb,twocolumn,10pt,a4paper,superscriptaddress,showpacs,floatfix]{revtex4}
\usepackage{amssymb,amsmath,bm}
\usepackage{setspace}
\usepackage[dvips]{graphicx}
\usepackage[dvips]{color}
\begin{document}

\title{Phase transitions and ordering of confined dipolar fluids}
\author{I. Szalai}
\affiliation{Institute of Physics, University of Pannonia, 
H-8201 Veszpr\'em, POBox 158, Hungary} 
\author{S. Dietrich}
\affiliation{Max-Planck-Institut f\"ur Metallforschung, Heisenbergstr. 3,
D-70569 Stuttgart, Germany}
\affiliation{
Institut f\"ur Theoretische und Angewandte Physik,
Universit\"at Stuttgart, Pfaffenwaldring 57, D-70569 Stuttgart, Germany}


\title{Phase transitions and ordering of confined dipolar fluids}

\begin{abstract}
We apply a modified mean-field density functional theory to determine the phase behavior of Stockmayer fluids in slitlike pores formed by two walls with identical substrate potentials. Based on the Carnahan-Starling equation of state, a fundamental-measure theory is employed to incorporate the effects of short-ranged hard sphere {\textendash} like correlations while 
the long-ranged contributions to the fluid interaction potential are treated perturbatively. The liquid-vapor, ferromagnetic liquid {\textendash} vapor, and ferromagnetic liquid {\textendash} isotropic liquid first-order phase separations are investigated. The local orientational structure of the anisotropic and inhomogeneous ferromagnetic liquid phase is also studied. We discuss how the phase diagrams are shifted and distorted upon varying the pore width.
\end{abstract}
\maketitle

\section{Introduction}
The structure and thermodynamics of confined dipolar fluids (such as molecular liquids, ferrofluids, and electrorheological fluids) have recently attracted considerable attention. Theoretical \cite{ch1,se1,kl1}, computer simulation \cite{we1,kl2,kl3,kl5,ha1,bo1}, and experimental \cite{ex1,ex2} studies have provided significant information about the orientational and spatial arrangement of dipolar particles in the vicinity of solid walls. The strongly anisotropic and long-ranged character of dipolar forces causes particular difficulties for the theoretical description of these systems. As long as one is not aiming for a quantitative description of a 
specific system but for general phenomena and trends, the so-called Stockmayer model has turned out to be rather useful \cite{gr1}. It considers spherical particles interacting with Lennard-Jones (LJ) pair potentials plus pointlike permanent dipoles at their centre. In this paper we adopt the magnetic language, assuming that the particles
carry magnetic dipole moments of strength $m$ because the main applications we have in mind are ferrofluids. (The results are identical for ferroelectric particles.)

Bulk dipolar fluids such as the Stockmayer model system may exhibit three distinct fluid phases: isotropic vapor, isotropic liquid, and ferromagnetic liquid \cite{gr1,wi1}.
(In the present study we focus on sufficiently high temperatures so that freezing is not a concern \cite{gr2}.) Accordingly the following first-order phase transitions can occur: isotropic vapor {\textendash} isotropic liquid, 
isotropic liquid {\textendash} ferromagnetic liquid,
and isotropic vapor {\textendash} ferromagnetic liquid. The phase diagrams exhibit the usual liquid-gas critical point, a triple point, a tricritical point, and a line of critical points corresponding to second-order phase transitions between the isotropic and the ferromagnetic liquid. For bulk dipolar fluids these phase transitions and critical points and lines have been detected in various theoretical studies \cite{gr1,wi1,gr2,kl4,sz1}. 

However, to the best of our knowledge there is no theoretical study which predicts the complete global phase diagrams and the structure of {\it confined} dipolar fluid phases.    
Recently Gramzow and Klapp \cite{kl1} have studied the phase behavior and orientational structure of Stockmayer fluids confined to slit pores. Within their modified mean-field (MMF) density functional theory (DFT) they considered only the subspace of spatially homogeneous local densities throughout the slit pore and focused on the orientational structure only. This allowed them to study the influence of the confinement on the phase behavior but neglecting the inevitable spatial inhomogeneities of the phases under study.
Our results will show that in the case of confined dipolar liquids, as for other fluids, there are strong spatial inhomogeneities in the vicinity of the walls. This inhomogeneous character of the liquid phases becomes more pronounced upon increasing of the density. These phenomena require a more sophisticated theory in order to describe the structure 
and the phase equilibria of confined dipolar systems.

In the present study we also use the modified mean-field (MMF) DFT approximation. However, for the description of the hard sphere reference system, instead of the homogeneous bulk Carnahan-Starling free energy expression, a fundamental measure theory free energy functional \cite{ro3} is applied in order to capture the short-ranged hard sphere {\textendash} like correlations. Our results for the 
confinement induced shifts and the distortions of the phase boundaries 
relative to the bulk ones are compared with the corresponding Kelvin equation.

The paper is organized as follows. In Sects. 2 and 3 the model and the extension of the MMF density functional theory for confined dipolar fluids are presented and developed, respectively. In Sec. 4 the calculation of the orientational order and the physical meaning of the order parameters are discussed. Sections 5 and 6 present the details of 
the Euler-Lagrange and of the Kelvin equations, respectively. The results of our calculations and a discussion are given in Sec. 7. Certain important computational details are summarized in the Appendices A, B, and C.

\section{Microscopic model}
\begin{widetext}
We study so-called hard core {\it S}tockmayer fluids which are characterized by the interaction potential
\begin{equation}
u_{S}(\mathbf{r}_{12},\omega_1,\omega_2)
=\left\{
      \begin{array}{lll}
     4\epsilon[(\sigma/r_{12})^{12}-(\sigma/r_{12})^{6}]
     -m^2D(\omega_{12},\omega_1,\omega_2)/r^3_{12} &, & r_{12} \geq \sigma\\
     \infty &, & r_{12} < \sigma,
        \end{array} 
        \right. \
\label{stm}
\end{equation}
\end{widetext}
where

\begin{eqnarray}
D(\omega_{12},\omega_1,\omega_2)\,\,\,\,\,\,\,\,\,\,\,\,\,\,\,\,\,\,\,
\,\,\,\,\,\,\,\,\,\,\,\,\,\,\,\,\,\,\,\,\,\,\,\nonumber\\
=3[\widehat{\mathbf{m}}_1(\omega_1)\cdot\widehat{\mathbf{r}}_{12}]
[\widehat{\mathbf{m}}_2(\omega_2)\cdot\widehat{\mathbf{r}}_{12}]
-[\widehat{\mathbf{m}}_1(\omega_1)\cdot\widehat{\mathbf{m}}_2(\omega_2)]
\label{D}
\end{eqnarray}
is a rotationally invariant function. 
In Eq. (\ref{stm}) the first term is the Lennard-Jones potential with length and energy 
parameters $\sigma$ and $\epsilon$, respectively. The second term in Eq. (\ref{stm}) together with Eq. (\ref{D}) describes the dipole-dipole interaction potential, where  particle 1 (2) with diameter $\sigma$ is located at $\mathbf{r}_1$ ($\mathbf{r}_2$) and carries at its center a point dipole moment of strength $m$ with an orientation 
given by the unit vector 
$\widehat{\mathbf{m}}_1(\omega_1)$ ($\widehat{\mathbf{m}}_2(\omega_2)$) with polar angles 
$\omega_1=(\theta_1,\phi_1)$ ($\omega_2=(\theta_2,\phi_2)$); 
$\mathbf{r}_{12}=\mathbf{r}_1-\mathbf{r}_2$ is the difference vector between the
centres of particle 1 and 2, $r_{12}=|\mathbf{r}_{12}|$, and 
$\widehat{\mathbf{r}}_{12}=\mathbf{r}_{12}/r_{12}$ is a unit vector with 
orientation $\omega_{12}=(\theta_{12},\phi_{12})$. In order to apply density functional theory this interaction potential is decomposed into a short-ranged repulsive part
\begin{equation}
u_r(r_{12})=u_{hs}(r_{12})=\left\{
        \begin{array}{lll}
         0 &, & r_{12} \geq \sigma \\
        \infty &, & r_{12} < \sigma 
        \end{array} 
        \right .,
\label{uhs}
\end{equation}
and into a long-ranged excess part
\begin{equation}
u_{exc}(\mathbf{r}_{12},\omega_1,\omega_2)=
\Theta(r_{12}-\sigma)u_S(\mathbf{r}_{12},\omega_1,\omega_2).
\label{ue}
\end{equation}
Thus the total pair potential is given by
\begin{equation}
u_S(\mathbf{r}_{12},\omega_1,\omega_2)=u_r(r_{12})+u_{exc}(\mathbf{r}_{12},\omega_1,\omega_2).
\label{S2}
\end{equation}
This decomposition lends itself to choose a hard-sphere reference system characterized by $u_r(r_{12})=u_{hs}(r_{12})$ for which reliable approximations for the free energy are known. 
The excess part $u_{exc}(\mathbf{r}_{12},\omega_1,\omega_2)$ will be treated 
perturbatively in an appropriate way.

For the fluid-wall interaction we consider two different potentials. In the case of 
a purely hard wall we use the hard-wall potential
\begin{equation}
u_{wh}(z)=\left\{
        \begin{array}{lll}
	0 &, &  \vert{z}\vert<(L-\sigma)/2\\
        \infty &, & \vert{z}\vert\geq(L-\sigma)/2,
        \end{array} 
        \right. \
\label{uhw}
\end{equation}
where $z$ is the coordinate normal to the walls and $L$ is the 
distance between the surfaces of the hard, parallel walls, which we call repulsive walls.
For the study of attractive walls we choose the external substrate potential
\begin{eqnarray}
u_{wa}(z)\,\,\,\,\,\,\,\,\,\,\,\,\,\,\,\,\,\,\,\,\,\,\,\,\,\,\,\,
\,\,\,\,\,\,\,\,\,\,\,\,\,\,\,\,\,\,\,\,\,\nonumber\\ 
=\left\{
        \begin{array}{lll}
	-\frac{2\pi}{3}\epsilon_w\left[(\frac{\sigma}{L/2+z})^3
	+(\frac{\sigma}{L/2-z})^3\right]
	 &, &  \vert{z}\vert<(L-\sigma)/2\\
        \infty &, & \vert{z}\vert\geq(L-\sigma)/2,
        \end{array} 
        \right. \nonumber \!\!\!\!\!\!\!\!\!\!\!\!\!\!\!\!\\
\label{usw}
\end{eqnarray}
where $\epsilon_w$ sets the energy scale. This equation can be obtained from the assumption that both walls consist of LJ particles interacting with the fluid particles of the same size also via the LJ potential with the energy scale $\epsilon_w$. Averaging the LJ interactions over all positions of the wall particles and approximating the averaged repulsive part of the LJ potential by a hard core potential one obtains the fluid particle {\textendash} wall interaction potential given by Eq. (\ref{usw}).

\section{Modified mean-field density functional theory for one-component fluids}

For a one-component nonuniform fluid configuration
$\hat{\rho}(\mathbf{r},\omega)$ denotes the number density of dipolar particles at a
point $\mathbf{r}=(x,y,z)$ with an orientation $\omega=(\theta,\phi)$ relative to 
a spatially fixed coordinate system.
The total number density of particles, independent of orientation, is given as
\begin{equation}
\hat\rho(\mathbf{r})=\int{d\omega\hat{\rho}(\mathbf{r},\omega)}=\int_{0}^{2\pi}
{d\phi}\int_{0}^{\pi}
{d\theta\sin(\theta)\hat{\rho}(\mathbf{r},\omega)}.
\end{equation} 
This allows one to split $\hat{\rho}(\mathbf{r},\omega)$ into the total number density $\hat\rho(\mathbf{r})$ and
a normalized space- and angle-dependent orientational distribution function
$\hat\alpha(\mathbf{r},\omega)$:
\begin{equation}
\hat{\rho}(\mathbf{r},\omega)=
\hat\rho(\mathbf{r})\hat\alpha(\mathbf{r},\omega),\ \ \ \ \ \ 
\int{d\omega\,\hat\alpha(\mathbf{r},\omega)}=1.
\label{norm}
\end{equation}
Within density functional theory \cite{ev1} the equilibrium density 
distribution $\rho(\mathbf{r},\omega;T,\mu)$ of an inhomogeneous fluid 
in the presence of an external potential $u_w$ minimizes the grand canonical potential
\begin{eqnarray}
\Omega[\{\hat\rho(\mathbf{r},\omega)\},T,\mu]\,\,\,\,\,\,\,\,\,\,\,\,\,\,\,\,\,\,\,\,\,\,\,\,\,\,\,\,\,\,\,\,\,\,\,\,\,\,\,\,\,\,\nonumber\\
=F[\{\hat\rho(\mathbf{r},\omega)\},T]
+\int{d^3rd\omega}\hat\rho(\mathbf{r},\omega)(u_w(\mathbf{r},\omega)-\mu)\,\,\,\,\,\,\,\,\,\,\nonumber\\
+\int{d^3r\,\kappa(\mathbf{r})\left(1-\int{d\omega\,{\hat\alpha}(\mathbf{r},\omega)}
\right)},\;\;\;\;\;\;
\label{omega}
\end{eqnarray}
and thus satisfies the Euler-Lagrange equation
\begin{equation}
\frac{\delta\Omega[\{\hat\rho(\mathbf{r},\omega)\},T,\mu]}{\delta\hat\rho(\mathbf{r},\omega)}
\Bigg|_{\hat\rho(\mathbf{r},\omega)=\rho(\mathbf{r},\omega;T,\mu)}=0.
\end{equation}
Here $F$ is the Helmholtz free energy density functional of the system 
and $\mu$ is the chemical potential. The last term in Eq. (\ref{omega}) takes into 
account the normalization condition (Eq. (\ref{norm})) for the orientational distribution 
function $\hat\alpha(\mathbf{r},\omega)$ by a Lagrange parameter $\kappa(\mathbf{r})$. 
Due to the two distinct contributions to the particle interaction 
potential (Eq. (\ref{S2})) the free energy functional $F$ decomposes into an 
ideal gas term $F^{id}$ and two corresponding parts:
\begin{eqnarray}
F[\{\hat\rho(\mathbf{r},\omega)\},T]=F^{id}[\{\hat\rho(\mathbf{r},\omega)\},T]
\nonumber\\
+F^{ref}[\{\hat\rho(\mathbf{r})\},T]+F^{exc}[\{\hat\rho(\mathbf{r},\omega)\},T],
\label{F}
\end{eqnarray}
where $F^{ref}$ is the reference and $F^{exc}$ is the excess free energy 
density functional.

In order to describe the planar wall - fluid interfaces, we consider a slab-shaped 
macroscopic system with the surfaces of the slab  parallel to the $x - y$ plane. 
The distance between the parallel surfaces is $L$. Thus, apart from spontaneous symmetry breaking like freezing, the equilibrium configuration of the system is 
inhomogeneous only in the $z$ direction and translationally invariant 
in the $x$ and $y$ directions. Under these circumstances the total number density 
$\rho(\mathbf{r})=\rho(z)$ is a function of $z$ only and the orientational profile 
$\alpha(\mathbf{r},\omega)=\alpha(z,\omega)$ depends on $z$ and the 
polar angle $\omega$.
In a coordinate system fixed in space the actual orientational distribution function
\begin{equation}
\alpha(z,\omega)=\sum_{l=0}^{\infty}\sum_{m=-l}^l\alpha_{lm}(z)Y_{lm}(\omega),
\label{al} 
\end{equation}
can be expanded in terms of the spherical harmonics $Y_{lm}(\omega)$ where 
$\alpha_{00}=1/\sqrt{4\pi}$ due to the normalization and 
$\alpha_{lm}^*=(-1)^m\alpha_{l\overline{m}}$ because $\alpha$ is real. 
(Here and in the following $\overline{m}\equiv-m$.) 
The expansion coefficients $\alpha_{lm}(z)$, which can be interpreted as 
orientational order parameters, are related to the full distribution function via
\begin{equation}
\alpha_{lm}(z)=\int{d\omega}\alpha(z,\omega)Y_{lm}^*(\omega). 
\label{alma}
\end{equation}
As presented below, for a slab-shaped system of transversal 
size $L$ one obtains explicit expressions for the three terms of the 
Helmholtz free energy density functional (see Eq. (\ref{F})).
\subsection{The free energy functional for the ideal gas}
For the slablike shape of the sample under consideration the ideal gas contribution 
has the form
\begin{eqnarray}
\!\!\!\!\!\!\!\!\!\!\!\!\!\!\!\!\!\!\!\!\!\!\!\!\!\!\!\!\!\!\!\!
F^{id}=\frac{A}{\beta}\int_{-L/2}^{L/2}{dz}\hat\rho(z)[\ln(\hat\rho(z)\Lambda^3)-1]\nonumber\\
+\frac{A}{\beta}\int_{-L/2}^{L/2}{dz}\hat\rho(z)\int{d\omega}\hat\alpha(z,\omega)
\ln[4\pi\hat\alpha(z,\omega)],
\end{eqnarray}
where $A$ is the lateral cross-sectional area of the system, $\Lambda$ is the de Broglie wavelength, 
and $\beta=1/(k_BT)$ is the inverse temperature.
\subsection{The reference free energy functional}
For the description of the hard sphere reference free energy functional we adopt the fundamental measure theory, which was initially proposed by Rosenfeld \cite{ro1,ro2}:
\begin{equation}
F^{ref}=\frac{A}{\beta}\int_{-L/2}^{L/2}{dz}\Phi(\{n_{\alpha}(z)\});
\end{equation}
the function $\Phi$ and the weighted densities $n_{\alpha}(z)$ are defined in Appendix A.
\subsection{The excess free energy functional}
We approximate the excess free energy functional in terms of the modified 
mean-field density functional theory \cite{te1,fr1,fr2}. Using Eq. (\ref{al}) for the 
present slablike geometry one obtains the following expression:
\begin{widetext}
\begin{eqnarray}
F^{exc}=\frac{1}{8\pi^2\beta}\int_{-L/2}^{L/2}{dz_1}\int_{-L/2}^{L/2}{dz_2}
\int_{A}{dx_1}{dy_1}\int_{A}{dx_2}{dy_2\;}\times\nonumber\\
\hat\rho(z_1)\hat\rho(z_2)\sum_{l_1,m_1}\sum_{l_2,m_2}
\hat\alpha_{l_1m_1}(z_1)\hat\alpha_{l_2m_2}(z_2)\times\nonumber\\
\int{d\omega_1}\int{d\omega_2}Y_{l_1m_1}(\omega_1)e^{-u_{ref}(r_{12})}
(1-e^{-u_{exc}(\mathbf{r}_{12},\omega_1,\omega_2)})Y_{l_2m_2}(\omega_2).
\label{fexc}
\end{eqnarray} 
\end{widetext}
Here and in the following the summations over $l_i,m_i$ are taken as
\begin{equation}
\sum_{l_i,m_i}...=\sum_{l_i=0}^{\infty}\sum_{m_i=-l_i}^{l_i}...\,\,\,\,\,.  
\end{equation}
In order to extract the explicit proportionality to the large cross-sectional area 
$A$ we introduce the sum and the difference 
of the lateral coordinates $(x_1,y_1)$ and $(x_2,y_2)$:
\begin{eqnarray}
x_{12}=x_1-x_2,\;\;\;\;y_{12}=y_1-y_2,\nonumber\\
x=\frac{x_1+x_2}{2},\;\;\;\;
y=\frac{y_1+y_2}{2}.
\end{eqnarray}
In terms of these new variables the integration of any function $f(r_{12})$ 
can be written as
\begin{eqnarray}
\int_A{dx_1dy_1}\int_A{dx_2dy_2}f(r_{12})\,\,\,\,\,\,\,\,\,\,\,\,\,\,\,\,\,\,\,\,\,\,\,\,\,\,\,\,\,\,\,\,\,\,\,\,\nonumber\\
=\int_A{dxdy}\int_A{dx_{12}dy_{12}}
f(r_{12})=A\int_A{dx_{12}dy_{12}}f(r_{12}).
\label{int}
\end{eqnarray}
In order to proceed we replace the rectangular coordinates $(x_{12},y_{12})$ 
by polar coordinates $(R_{12},\phi_{12})$ and take into account Eqs. (\ref{fexc}) 
and (\ref{int}) for the $\textit{exc}$ess free energy functional:
\begin{eqnarray}
\frac{F^{exc}}{A}=\int_{-L/2}^{L/2}{dz_1}\int_{-L/2}^{L/2}{dz_2}\;
\hat\rho(z_1)\hat\rho(z_2)\times\nonumber\\
\sum_{l_1,m_1}\sum_{l_2,m_2}\hat\alpha_{l_1m_1}(z_1)
\hat\alpha_{l_2m_2}(z_2)u_{l_1m_1l_2m_2}(z_1-z_2),
\end{eqnarray}
where
\begin{eqnarray}
u_{l_1m_1l_2m_2}(z_1-z_2)
=-\frac{1}{8\pi^2\beta}\int_0^{\infty}{dR_{12}R_{12}}\Theta(r_{12}-
\sigma)\times\nonumber \!\!\!\!\\
\int_0^{2\pi}{d\phi_{12}}
\int{d\omega_1}\int{d\omega_2}Y_{l_1m_1}(\omega_1)
f_M(\mathbf{r}_{12},\omega_1,\omega_2)Y_{l_2m_2}(\omega_2).\nonumber 
\!\!\!\!\!\!\!\!\!\!\\
\label{ukl}
\end{eqnarray}
In Eq. (\ref{ukl}) $f_M$ denotes the Mayer function of the excess
potential:
\begin{equation}
f_M(\mathbf{r}_{12},\omega_1,\omega_2)=
e^{-{\beta}u_{exc}(\mathbf{r}_{12},\omega_1,\omega_2)}-1,
\label{mayer}
\end{equation}
where in accordance with the planar polar coordinates 
used for the slab-geometry
\begin{equation}
r_{12}=\sqrt{R_{12}^2+(z_1-z_2)^2}.
\end{equation}
Via the vector $\widehat{\mathbf{r}}_{12}$ the rotationally invariant function 
$D(\omega_1,\omega_2,\omega_{12})$ (Eq. (\ref{D})) depends on $\vartheta_{12}$ as well. 
Therefore in Eq. (\ref{ukl}) the substitutions
\begin{eqnarray}
\sin\vartheta_{12}=\frac{R_{12}}{\sqrt{R_{12}^2+(z_1-z_2)^2}},\nonumber\\
\cos\vartheta_{12}=\frac{z_1-z_2}{\sqrt{R_{12}^2+(z_1-z_2)^2}}
\end{eqnarray}
have to be carried out prior to the integration over the variable $R_{12}$. 
In our calculation we have expanded the exponential term in Eq. (\ref{mayer}), 
containing the dipole-dipole interaction, into a Taylor series:
\begin{equation}
f_M=-1+e^{-{\beta}u_{LJ}(r_{12})}\left[1
+\frac{{\beta}m^2}{(r_{12})^3}D 
+\frac{1}{2\;!}\frac{{\beta}^2m^4}{(r_{12})^6}D^2...\right].
\label{dtaylor}
\end{equation}
In accordance with Eqs. (\ref{ukl}) and (\ref{dtaylor}) $u_{l_1m_1l_2m_2}$ is given as a Taylor
series, too:
\begin{eqnarray}
u_{l_1m_1l_2m_2}=u_{l_1m_1l_2m_2}^{(0)}\nonumber\\
+u_{l_1m_1l_2m_2}^{(1)}\times(m^2)^1
+u_{l_1m_1l_2m_2}^{(2)}\times(m^2)^2...,
\label{ukk}
\end{eqnarray}
with the coefficients
\begin{eqnarray}
u^{(i)}_{l_1m_1l_2m_2}(z)=-\frac{\beta^{\;i-1}}{2i\;!}\int_0^{\infty}
dR_{12}R_{12}\nonumber\\
\times\Theta(R_{12}^2+z^2-\sigma^2)
\frac{\exp{\left[-{\beta}u_{LJ}\left(\sqrt{R_{12}^2+z^2}\right)\right]}-\delta_{i0}}{(R_{12}^2+z^2)^{3i/2}}\nonumber\\
{\times}A_{l_1m_1l_2m_2}^{(i)}\left(\frac{z}{\sqrt{R_{12}^2+z^2}}\right).
\label{ukl2}
\end{eqnarray}
In Eq. (\ref{ukl2}) the coefficients $A^{(i)}_{l_1m_1l_2m_2}$ can be expressed as
\begin{eqnarray}
A_{l_1m_1l_2m_2}^{(i)}(\cos\theta_{12})\nonumber\\
=\int_{0}^{2\pi}{d\phi_{12}}
\int{d\omega_1}
\int{d\omega_2}
Y_{l_1m_1}(\omega_1)\nonumber\\
{\times}[D(\omega_{12},\omega_1,\omega_2)]^iY_{l_2m_2}(\omega_2).
\label{aikl}
\end{eqnarray}
In Appendix B we present the nonzero coefficients $A_{l_1m_1l_2m_2}^{(i)}$ 
for $i,l_1,l_2=0,1,2$. In the bulk limit the expressions for $u_{l_1m_1l_2m_2}$ 
reduce to the corresponding ones in Ref. \cite{gr1}:
\begin{eqnarray}
\lim_{L\rightarrow{\infty}}\int_{-L/2}^{L/2}{dz\,u^{(i)}_{l_1m_1l_2m_2}(z)}
\nonumber\\=\pi\sqrt{(2l_1+1)(2l_2+1)}\delta_{l_1l_2}
\delta_{m_10}\delta_{m_20}u^{(i)}_{l_1},
\end{eqnarray}  
where the coefficients $u^{(i)}_{l_1}$, for $l_1=0,1,2,3,4$, are given by Eqs. (4.9)-(4.13) in 
Ref. \cite{gr1}. It is important to note that among the coefficients $u_{l_1m_1l_2m_2}$ 
there are terms which are nonzero in a slablike geometry,
but which vanish in the bulk limit.
\section{Magnetization and order parameters}
Focusing on the structure of confined dipolar fluids, here we are particularly interested in the occurrence of a spontaneous magnetization. In the slab geometry, which can be considered as the limiting case of an oblate ellipsoidal sample with vanishing aspect ratio, the demagnetization field vanishes if the fluid sample is magnetized along a spontaneously chosen direction within the $xy$ plane. For such a type of magnetization the formation of various domains can be excluded. Therefore the confined fluid with the magnetization within the $xy$ plane is comparable to a single domain bulk fluid in a needlelike volume with longitudinal magnetization. The free energies of these fluid samples can be mapped onto each other \cite{gr1}.  
The definition of the local magnetization is
\begin{equation}
\mathbf{M}(z)=m\rho(z)\int{d\omega}\alpha(z,\omega)\widehat{\mathbf{m}}(\omega). 
\end{equation}
By inserting the expansion in terms of the spherical harmonics and performing the integration 
we obtain
\begin{equation}
\mathbf{M}(z) =\left(\frac{4\pi}{3}\right)^{1/2}m\rho(z)\left\{
        \begin{array}{lll}
        -\sqrt{2}{Re}{[\alpha_{11}(z)]} \\
	 +\sqrt{2}{Im}{[\alpha_{11}(z)]} \\
	\alpha_{10}(z)
        \end{array}
	\right\}. 
\label{us}
\end{equation}
For $M_z=0$, i.e., $\alpha_{10}(z)=0$, the magnetization within the $xy$ plane is 
measured by the order parameter
\begin{eqnarray}
\alpha_{xy}(z)=\left(\frac{8\pi}{3}\right)^{1/2}
\left((Re{[\alpha_{11}(z)]})^2+(Im{[\alpha_{11}(z)]})^2\right)^{1/2}\nonumber\\
=\left(\frac{8\pi}{3}\right)^{1/2}\vert\alpha_{11}(z)\vert,\;\;\;\;\;\;\;\;\;\;
\label{axy}
\end{eqnarray}
so that the modulus $M_{xy}$ of the in-plane magnetization is
\begin{equation}
M_{xy}(z)=m\rho(z)\alpha_{xy}(z).
\label{mxy} 
\end{equation}
The higher coefficients of the orientational distribution function $\alpha(z,\omega)$ (Eq. (\ref{al})) may also be considered as local orientational order parameters. 
Nonzero coefficients for $l=2$ indicate a certain type of ordering of the dipoles. 
An especially interesting quantity is the 
order parameter $\alpha_{20}(z)$ that describes the orientation of dipole-axes relative 
to the $z$ axis. This is related to the $zz$ component of the local quadrupole tensor as
\begin{equation}
Q_{zz}(z)=\int{d\omega\alpha(z,\omega)P_2(\cos\theta)}=\left(\frac{4\pi}{5}\right)^{1/2}
\alpha_{20}(z),
\end{equation}
where $P_2$ is the second order Legendre polynomial. $Q_{zz}=1$ corresponds to the perfect 
ordering of dipole axes normal to the walls while $Q_{zz}=-1/2$ indicates the perfect 
ordering along directions in the $xy$ plane. Randomly oriented dipoles correspond to $Q_{zz}=0$.
\\


\section{Euler-Lagrange equation} The equilibrium configuration characterized by $\rho(z)$ and $\alpha(z,\omega)$ for given values of $T$ and $\mu$ follows from minimizing 
the total grand canonical functional with respect to $\hat\rho(z)$ and the function 
$\hat\alpha(z,\omega)$:
\begin{widetext}
\begin{eqnarray}
\frac{\beta}{A}\frac{\delta\Omega}{\delta\hat\rho(z)}
\Bigg|_{\hat\rho=\rho,\,\,\hat\alpha=\alpha}=\ln\rho(z)
+\beta\mu^{ref}_{hs}[{\{\rho(z)}\}]+\int{d\omega\,{\alpha}(z,\omega)}
\ln[4\pi{\alpha}(z,\omega)]\nonumber\\
\!\!\!\!\!\!\!\!\!\!\!\!\!\!\!\!\!\!\!\!\!\!\!\!\!\!\!\!\!\!\!
+2\beta\sum_{l_1,m_1}\alpha_{l_1m_1}(z)\sum_{l_2,m_2}
\int_{-L/2}^{L/2}{dz'}\rho(z')u_{l_1m_1l_2m_2}(z-z')\alpha_{l_2m_2}(z')
-\beta\mu+\beta{u_w(z)}=0
\label{rhoder}
\end{eqnarray}
and
\begin{eqnarray}
\frac{\beta}{A}\frac{\delta\Omega}{\delta\hat\alpha(z,\omega)}
\Bigg|_{\hat\rho=\rho,\,\,\hat\alpha=\alpha}=\rho(z)\
(1+\ln[4\pi{\alpha}(z,\omega)])-\beta\kappa(z)\nonumber\\
\!\!\!\!\!\!\!\!\!\!\!\!\!\!\
+2\beta\rho(z)\int_{-L/2}^{L/2}{dz'}\rho(z')
\sum_{l_1,m_1}\alpha_{l_1m_1}(z')\sum_{l_2,m_2}u_{l_1m_1l_2m_2}(z-z')
Y_{l_2m_2}^*(\omega)=0,
\label{alpder}
\end{eqnarray}
\end{widetext}
where $\mu^{ref}_{hs}[{\{\hat\rho(z)}\}]=\frac{\delta F^{ref}_{hs}[{\{\hat\rho(z)}\}]}{\delta\hat\rho(z)}$ 
is the chemical potential functional (first order direct correlation function) of the 
hard sphere reference system (see Appendix A).
The corresponding set of integral equations is discussed in Appendix C.
As solutions of Eqs. (\ref{rhoder}) and (\ref{alpder}) $\rho(z;T,\mu)$ and $\alpha(z,\omega;T,\mu)$ become functions of $T$ and $\mu$.
\section{Phase equilibria and Kelvin equation}
The phase coexistence curves $\mu(T)$ and the coexisting densities and orientational order parameters follow from requiring the equality of the grand potentials for the coexisting phases $I$ and $II$:
\begin{eqnarray}
\Omega[\{{\rho^{(I)}(z;T,\mu),\alpha^{(I)}(z,\omega;T,\mu)}\},T,\mu]=\nonumber\\
\Omega[\{{\rho^{(II)}(z;T,\mu),\alpha^{(II)}(z,\omega;T,\mu)}\},T,\mu].
\label{equil}
\end{eqnarray}
The functions $\rho^{(i)}(z;T,\mu)$ and $\alpha^{(i)}(z,\omega;T,\mu)$ 
$(i=I,II)$ denote the corresponding equilibrium density and orientational distribution functions obtained from Eqs. (\ref{rhoder}) and (\ref{alpder}). Equation (\ref{equil}) requires simultaneous solutions of the Euler-Lagrange equations (\ref{rhoder}) 
and (\ref{alpder}) for a wide range of temperatures and 
chemical potentials. As mentioned in Sec. 1, we consider three kinds of two-phase equilibria: isotropic liquid {\textendash} isotropic vapor, isotropic vapor {\textendash} ferromagnetic liquid, and isotropic liquid {\textendash} ferromagnetic liquid.
In order to estimate in leading order of $1/L$ the location of the phase coexistence curves, the Kelvin equation (see, e.g. Ref. \cite{di1} and references therein) is used to obtain the chemical potential difference between the bulk phase points and the corresponding ones for slabs of thickness $L$ at a given temperature and for two identical confining walls:
\begin{equation}
\Delta\mu=
\mu(L)-\mu(\infty)=\frac{2}{L}\frac{\gamma_{wI}-\gamma_{wII}}{\rho_I-\rho_{II}}=
-\frac{2}{L}\frac{\gamma_{I,II}\cos\Theta_I}{\rho_I-\rho_{II}}
\label{keq}
\end{equation}
where $\rho_i$ ($i=I,II$) are the bulk densities of the coexisting phases, $\gamma_{wi}$ are the corresponding wall-fluid interfacial tensions, $\gamma_{I,II}$ is the surface tension between the coexisting fluid phases $I$ and $II$, and $\Theta_I=\arccos[(\gamma_{wII}-\gamma_{wI})/\gamma_{I,II}]$ is the contact angle of a drop of fluid $I$ with the wall (Young's equation). In the slab-shaped system, for a fluid which is in contact with two identical walls, the surface tension $\gamma_{wi}$ follows from the equilibrium density $\rho(z;T,\mu)$ and oriental distribution $\alpha(z,\omega;T,\mu)$ profiles as
\begin{equation}
\gamma_{wi}=\frac{1}{2}\lim_{L\rightarrow\infty}
\left(Lp+\Omega_L[\{\rho(z;T,\mu),\alpha(z,\omega;T,\mu)\},T,{\mu}]\right),
\label{ten}
\end{equation}
where $p$ is the bulk pressure and $\Omega_L$ is the grand potential per cross-sectional area A of the slablike system of transversal size $L$.

\section{Results and discussion}
In the following we shall use dimensionless quantities: $T^*=k_BT/\epsilon$ as reduced temperature, $\mu^*=(\mu-k_BT\ln(\Lambda^3/\sigma^3))/\epsilon$ as reduced chemical potential, $\bar{\rho}^*=\frac{\sigma^3}{L}\int_{-L/2}^{L/2}{dz}\rho(z)$ as reduced averaged density in the pore, which in a bulk phase equals the usual reduced density, 
$m^*=m/(\epsilon\sigma^3)^{1/2}$ as the reduced dipole moment, and $\epsilon_w^*=\epsilon_w/\epsilon$ for describing attractive walls (Eq. (\ref{usw})).

For calculating the expansion of the dipole-dipole interaction Mayer function we used a second order approximation. This choice restricts the expansion coefficients 
$\alpha_{lm}(z)$ of the orientational distribution function to those with $l\leq 2$ and thus $\arrowvert m\arrowvert \leq 2$. 
\subsection{Phase diagrams}
As a first step, on the basis of the corresponding bulk theory we have determined the bulk phase diagrams of Stockmayer fluids for different dipole moments. For comparison  and as a reference the corresponding bulk phase diagram is depicted in all figures discussing the phase diagrams in slabs. We note that the bulk phase diagrams presented here differ numerically from those in Ref. \cite{gr1} because there a fourth-order expansion with respect to the corresponding Mayer functions has been used as well as a temperature dependent hard sphere diameter for the reference system. Extending our present analysis of spatially inhomogeneous system also to fourth order in that expansion is in principle possible but requires unreasonably large technical efforts. The bulk phase diagrams within the present approximation exhibit only minor quantitative differences from those in Ref. \cite{gr1}.

For the dipole moment $m^*=1.5$ and the wall distance $L/\sigma=10$ Fig. 1 presents our numerical results for the phase diagrams in the chemical potential {\textendash} temperature and the density {\textendash} temperature planes. Figure 1 compares the bulk phase diagrams with those of the slab for repulsive or attractive walls. For describing the attractive walls (Eq.(\ref{usw})) here and in the following we choose $\epsilon_w^*=1/2$. Figure 1(a) shows that below the triple point temperature $T_{TP}$ the isotropic vapor coexists with the ferromagnetic liquid. The confinement by repulsive walls shifts the chemical potential {\textendash} temperature phase diagrams towards higher chemical potentials relative to the bulk data. In the case of attractive walls the shift is smaller and in the opposite direction. This figure also shows that for the chosen parameters the Kelvin equation provides a good estimate of these shifts. Between the temperatures $T_{TP}$ and $T_{CP}$ of the triple point temperature and of the liquid-vapor critical point, respectively, there are three possible phases: the isotropic vapor, the isotropic liquid, and the ferromagnetic fluid. The first-order phase transition between the isotropic liquid and ferromagnetic fluid turns into a second-order phase transition above the tricritical temperature $T_{TCP}$. (For large $\mu^*$ and at high $\rho^*$ orientationally disordered or ferromagnetic solid phases appear \cite{gr2}, which are not considered here.) 
\begin{figure}[t]
\includegraphics*[width=15.5pc]{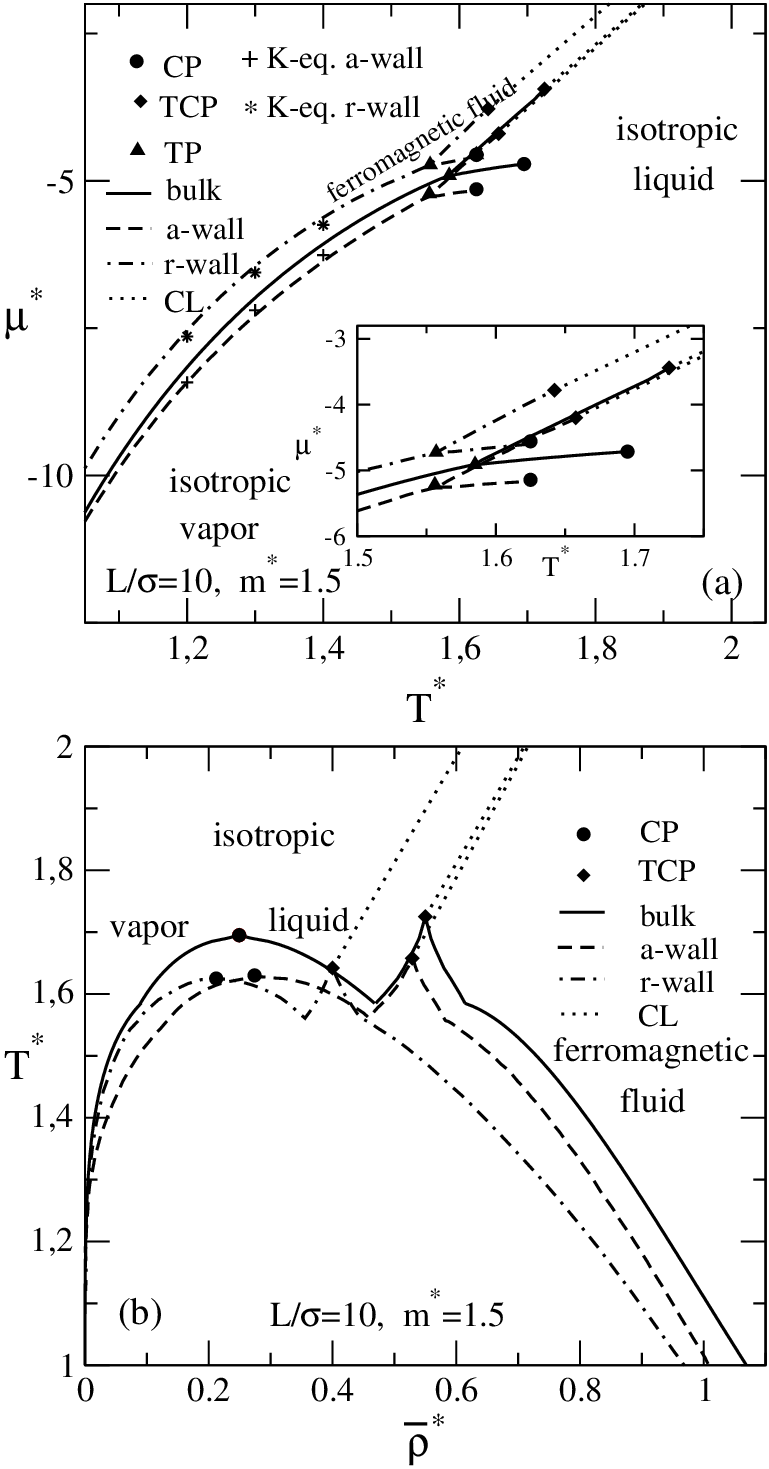}
\begin{spacing}{1.5}
\caption{\footnotesize Phase diagrams of bulk and confined Stockmayer fluids 
for $m^*=1.5$ and $L/\sigma=10$ in (a) the chemical potential {\textendash} temperature and (b) in the averaged density {\textendash} temperature plane; $\bar{\rho}^*=\frac{\sigma^3}{L}\int_{-L/2}^{L/2}dz\rho(z)$ is the spatially averaged overall number density. In (a) solid (bulk phase), dashed 
({\it a}ttractive wall, $\epsilon_w^*=1/2$), and dash-dotted ({\it r}epulsive wall) lines are lines of first-order phase transitions. In (b) these curves enclose two-phase regions. 
The dotted lines CL denote lines of second-order 
phase transitions. CP, TCP, and TP denote the liquid-vapor critical, 
the tricritical, and the triple point, respectively. Three-phase coexistence at TP is associated with and visible by breaks in slope of the corresponding first-order phase boundaries.  The symbols $+$ and $*$ indicate phase equilibrium points predicted by the corresponding {\it K}elvin equation for {\it a}ttractive and {\it r}epulsive walls, respectively.}
\end{spacing}
\end{figure}
\begin{figure}[t]
\includegraphics*[width=15.5pc]{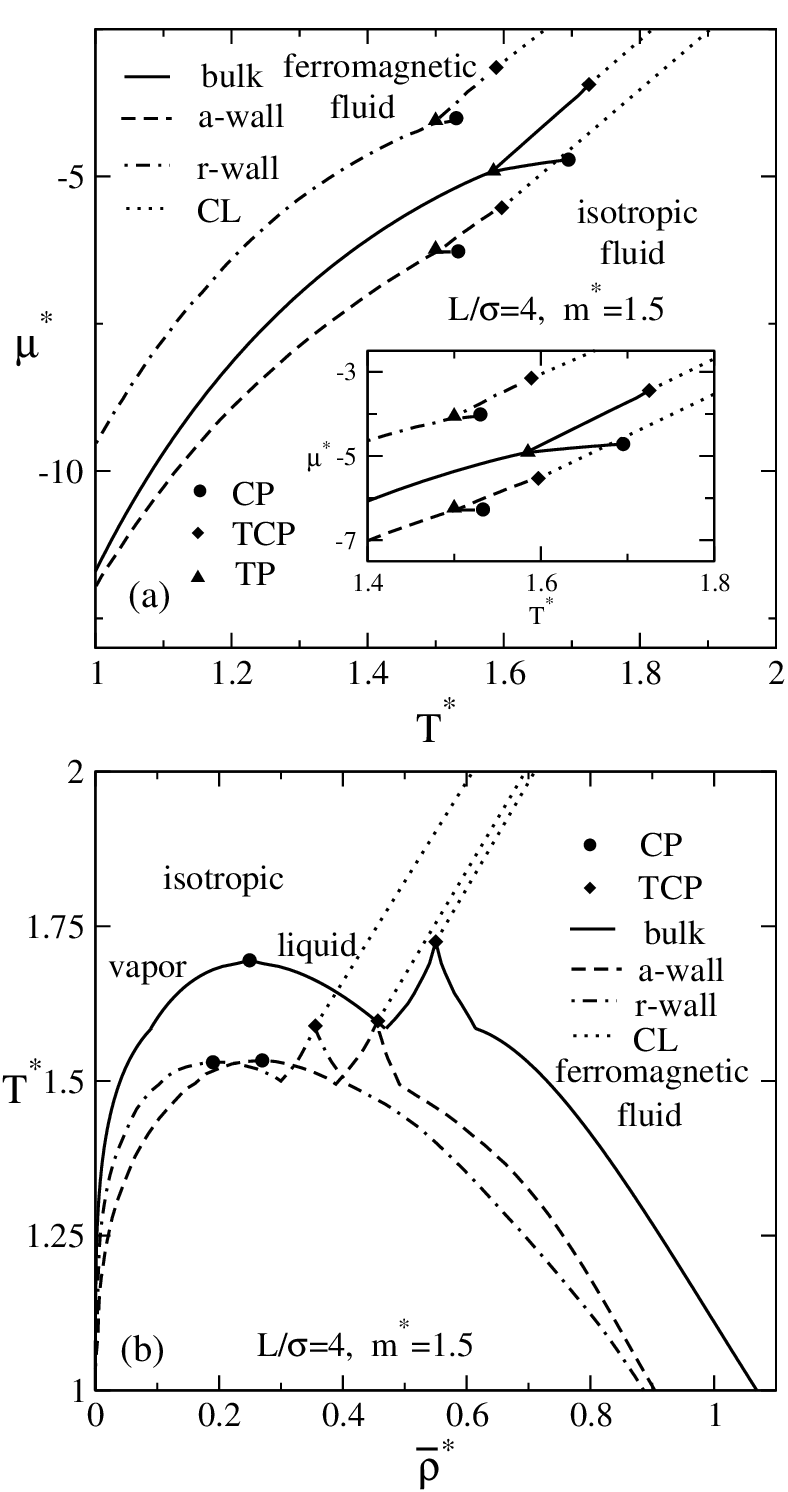}
\begin{spacing}{1.5}
\caption{\footnotesize Phase diagrams of bulk and confined Stockmayer fluids (as in Fig. 1) for $m^*=1.5$, $L/\sigma=4$, and $\epsilon_w^*=1/2$ in (a) the chemical potential {\textendash} temperature and (b) averaged density {\textendash} temperature plane.}
\end{spacing}
\end{figure}
We find that, compared with the bulk, the temperature range for liquid-vapor and liquid {\textendash} ferromagnetic fluid coexistence is narrowed by the confinement, both for repulsive and attractive walls. According to Fig. 1(b) at low temperatures ($T<T_{TP}$) the high density ferromagnetic fluid coexists with the low density isotropic vapor. Above the triple point temperature $T_{TP}$, at low and medium densities two isotropic fluids coexist, becoming identical above the liquid-vapor critical temperature $T_{CP}$. At higher densities, the isotropic liquid (with the lower density) and the ferromagnetic fluid (with the higher density) are separated by first-order phase transitions which turn into second-order phase transitions above the tricritical temperature $T_{TCP}$. The critical line CL of second-order phase transitions divides the liquidlike thermodynamic states into an isotropic liquid phase and into a ferromagnetic fluid phase. In comparison with the bulk coexistence curves the shift of the first-order coexistence region to lower densities is more significant for repulsive than for attractive walls. The tricritical liquid densities are lowered by the confinement in comparison with the corresponding bulk ones whereas the tricritical vapor densities are increased. The liquid-vapor critical 
density is lowered due to the confinement by repulsive walls and slightly increased in the case of attractive walls. At a given temperature both confinements promote the formation of the ferromagnetic phase at lower densities. These findings are in qualitative agreement with the results of Gramzow and Klapp \cite{kl1}, but the shift of the coexistence curves of confined systems relative to the bulk ones is more significant than predicted by their homogeneous local density theory.
\begin{figure}[t]
\includegraphics*[width=15.5pc]{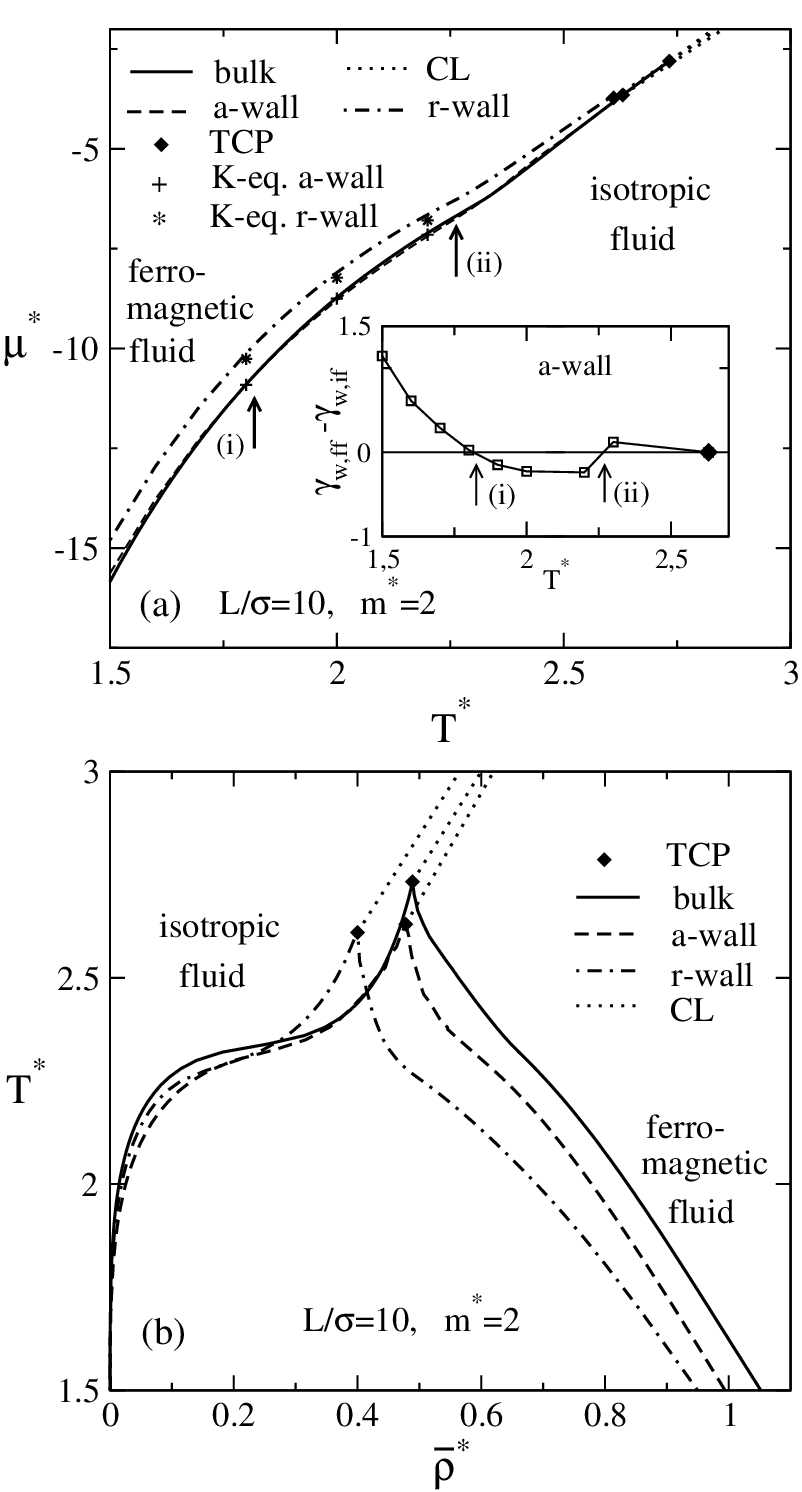}
\begin{spacing}{1.5}
\caption{\footnotesize Phase diagrams of bulk and confined Stockmayer fluids 
for $m^*=2$, $L/\sigma=10$, and $\epsilon_w^*=1/2$ in (a) the chemical potential {\textendash} temperature 
and (b) averaged density {\textendash} temperature plane. The line codes and the meaning of the symbols are the same as in Figs. 1 and 2. For this dipole strength the liquid-vapor critical point has disappeared both in the bulk and in the confined system. In (a) the attractive wall confinement barely shifts the phase boundary; nonetheless this shifted boundary crosses the bulk one twice. Within the Kelvin equation for the shift these crossings correspond to zeroes (i) and (ii) of the independently calculated surface tension difference $\gamma_{w,ff}-\gamma_{w,if}$ (see the inset) which correspond to a contact angle $\Theta_{ff}=90^{\circ}$ (see the main text).}
\end{spacing}
\end{figure}
As shown in Fig. 2, a stronger confinement ($L/\sigma=4$) does not change the topology of the phase diagram. However, in this case the shift of the phase boundaries relative to the bulk ones are more pronounced. The temperature interval $T_{CP}-T_{TP}$, within which liquid-vapor coexistence is thermodynamically stable, becomes narrower with increasing confinement. All critical temperatures ($T_{CP}$, $T_{TCP}, T_{TP}$) are lowered upon increasing the confinement. (The lowering of the liquid-vapor critical temperature for Stockmayer fluids with increasing confinement is in a qualitative aggrement with the Gibbs ensemble Monte Carlo simulation results of Richardi {\it et al}  \cite{we1}.) For attractive walls the critical density of the liquid-vapor equilibria does not change significantly whereas it is reduced for repulsive walls. Comparing our results for $m^*=1.5$ and $L/\sigma=4$ with those of Gramzow and Klapp \cite{kl1} we note that, in contrast to their findings, in the chemical potential {\textendash} temperature plane the first-order phase boundaries of the bulk and of the confined systems do not intersect.

For $L/\sigma=10$, increasing the dipole moment from $m^*=1.5$ to $m^*=2$ changes the corresponding phase diagrams shown in Fig. 1 considerably and leads to the phase diagram displayed in Fig 3. This increase of the dipole moment causes the disappearance of the liquid-vapor transitions for both the bulk and the confined systems. According to Fig. 3(a) for this dipole moment the phase boundaries of the bulk and of the confined system with attractive {\it w}alls do intersect. On the basis of the Kelvin equation (Eq. (\ref{keq})) this multiple crossing of the 
{\it f}erromagnetic {\it f}luid {\textendash} {\it i}sotropic {\it f}luid ($I=ff$, $II=if$) phase boundaries for the bulk and for the confined system implies a non-monotonic temperature dependence of the surface tension difference $\gamma_{w,ff}-\gamma_{w,if}$ and vice versa; according to the full line in Fig. 3(b) $\rho_{ff}-\rho_{if}>0$ varies monotonically as function of temperature. (For smaller dipole moments a similar crossing behavior has been found by Gramzow and Klapp \cite{kl1}.) 
The inset in Fig. 3(a) shows the independently calculated (Eq. (\ref{ten})) surface tension difference and confirms that the aforementioned crossings of the bulk and film phase boundaries do coincide with the changes of sign of $\gamma_{w,ff}-\gamma_{w,if}$ at $T_{(i)}$ and $T_{(ii)}$.
Due to the second equation in Eq. (\ref{keq}) the Kelvin equation turns into
\begin{equation}
\Delta\mu=
-\frac{2}{L}\frac{\gamma_{ff,if}\cos\Theta_{ff}}{\rho_{ff}-\rho_{if}}.
\end{equation}
Since $\gamma_{ff,if}$ is expected to decrease monotonically upon increasing temperature, these crossings amount to a non-monotonic temperature dependence of the contact angle of the ferromagnetic fluid around $90^{\circ}$. (For $\Theta_{ff}=90^{\circ}$ the leading behavior of $\Delta\mu$ is given by the contribution $\sim{1/L^2}$.)

In the case of confinement with repulsive walls the ferromagnetic fluid {\textendash} isotropic fluid phase boundary is generally shifted to higher values of the chemical potential as compared with its bulk counterpart. Confinement lowers the tricritical temperatures $T_{TCP}$ and there is no significant difference between the tricritical temperatures for systems with attractive and repulsive walls.
As for $m^*=1.5$ (Fig. 1(a)), also for $m^*=2$ and $L/\sigma=10$ the estimates obtained from the Kelvin equation are in good agreement with the full phase equilibrium calculations. Figure 3(b) shows how the two-phase region shrinks upon confinement. If the liquid-vapor critical point has disappeared in the bulk phase diagram, this holds also for the confined system, independent of the character of the wall.
\begin{figure}[t]
\includegraphics*[width=15.5pc]{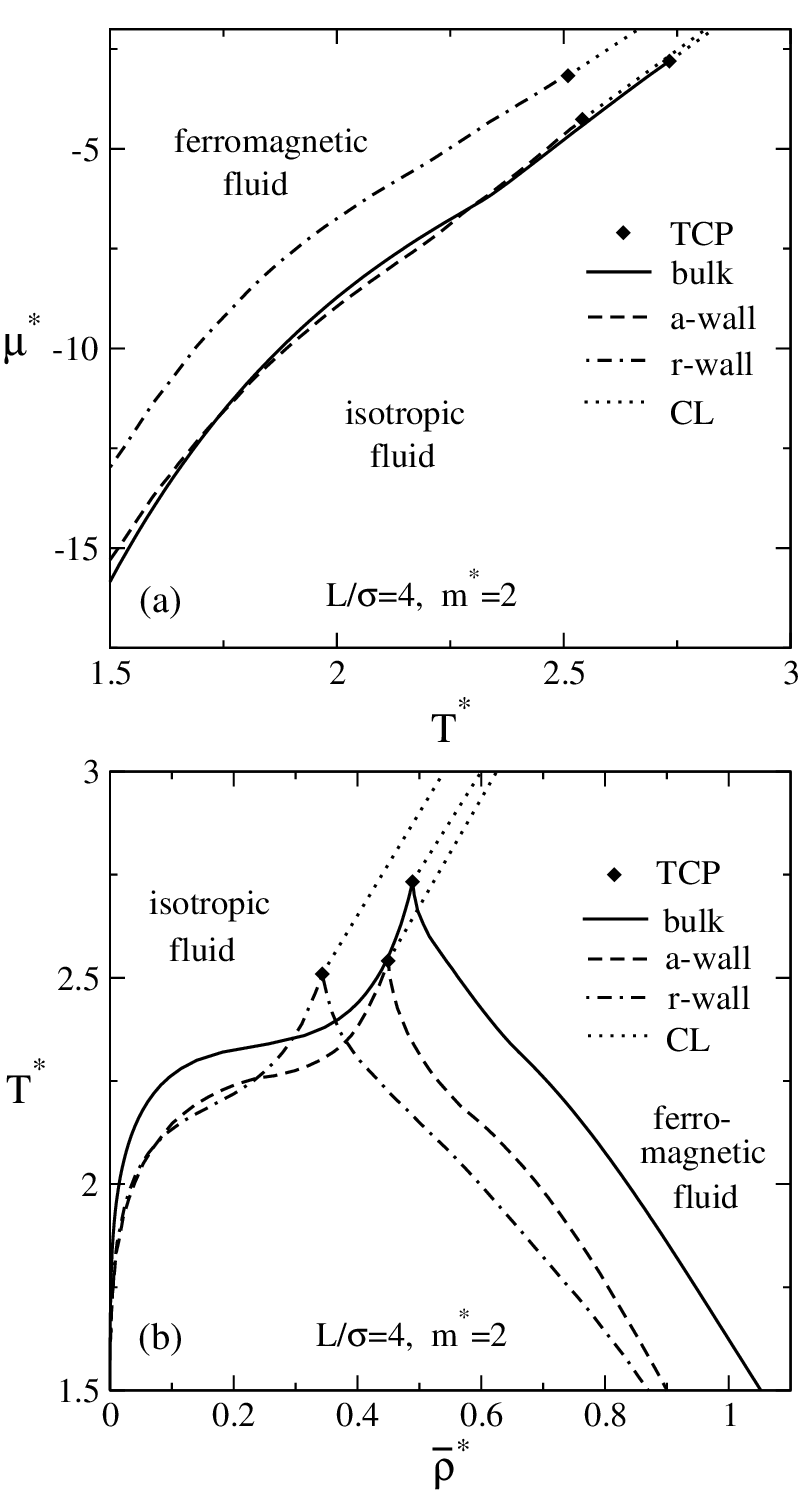}
\begin{spacing}{1.5}
\caption{\footnotesize Phase diagrams of bulk and confined Stockmayer fluids 
for $m^*=2$, $L/\sigma=4$, and $\epsilon_w^*=1/2$ in (a) the chemical potential {\textendash} temperature 
and (b) averaged density {\textendash} temperature plane. The line codes and the meaning of the symbols are the same as in Figs. 1-3. In (a) the crossings of the shifted phase boundary for the attractive wall confinement with the bulk phase boundary is more clearly visible than for $L/\sigma=10$ shown in Fig. 3(a).}
\end{spacing}
\end{figure}
For $m^*=2$, upon decreasing the wall separation from $L/\sigma=10$ to $L/\sigma=4$ the topology of the phase diagrams does not change (compare Figs. 3 and 4). Figure 4(a) shows that for repulsive walls the stronger confinement leads to a larger shift of the phase boundary to higher values of the chemical potentials. For attractive walls this shift remains much smaller even for this narrow slab 
with $L/\sigma=4$. Different from the case $m^*=1.5$ (see Fig. 2(a)), for $m^*=2$ two crossings of the bulk coexistence curve with the phase boundary for attractive walls are clearly visible. Their interpretation in terms of the Kelvin equation is less valid than for Fig. 3(a) because due to the small value $L/\sigma=4$ subdominant terms in the expression for $\Delta\mu$ become relevant.
Figure 4(b) shows that enhancing the confinement causes a further shrinking of the two-phase coexistence region. The tricritical temperatures $T_{TCP}$ and densities $\rho_{TCP}$ decrease upon reducing the wall separation. This means that for smaller distances between the walls the ferromagnetic fluid {\textendash} isotropic fluid phase coexistence occurs within a narrower range of the thermodynamic variables.

In a recent publication Trasca and Klapp \cite{kl5} have studied inter alia the 
second-order phase transition of strongly coupled dipolar fluids confined to narrow slit pores. They performed Monte Carlo simulations using purely repulsive wall confinements and the dipolar soft sphere model for the particles. In confined systems, they have found that, at a given temperature, with the decrease of the wall separation the paramagnetic-ferromagnetic phase transition density (averaged across the slit pore) decreases relative to the corresponding bulk one. They found that the direction of this shift is inconsistent with their very simple mean-field theoretical predictions (see Eq. (3a) in Ref. \cite{kl5}). Analyzing the shifts of the second-order transition lines for the repulsive wall confinements (see Figs. 3(b) and 4(b)) we find that within the framework of our DFT at a given temperature the critical averaged density decreases upon the decrease of the wall separation, which is in qualitative aggrement with the simulation findings of Trasca and Klapp \cite{kl5}. (There is no possibility for a quantitative comparison because these Monte Carlo simulations were carried out for a reduced dipole moment $m^*=3$; for such large values of $m^*$ the quantitative reliability of the present DFT would be reduced anyhow.)

\subsection{Structural properties}
In the following the structural properties of the coexisting ferromagnetic fluid and isotropic gas phases are discussed. Figure 5 shows the corresponding results for reduced dipole moment $m^*=1.5$ and wall separation $L/\sigma=10$ at the reduced temperature $T^*=1.15$. At this low temperature the ferromagnetic fluid can be denoted as a liquid (see Fig. 1(b)). (As a hint we note that for a given temperature the chemical potential at two-phase coexistence is different for attractive and repulsive walls, see Fig 1(a).) 
Figure 5(a) shows that the confined ferromagnetic liquid phase is strongly structured both for attractive and repulsive walls. In the case of confinement by attractive walls the contact value $\rho^*_w\simeq19.8$ of the reduced density (i.e., at $|{z}|=(L-\sigma)/2$) is very high (not displayed on the scale of the figure) which refers to a strong adsorption on the walls. 
\begin{figure}[t]
\includegraphics*[width=20pc]{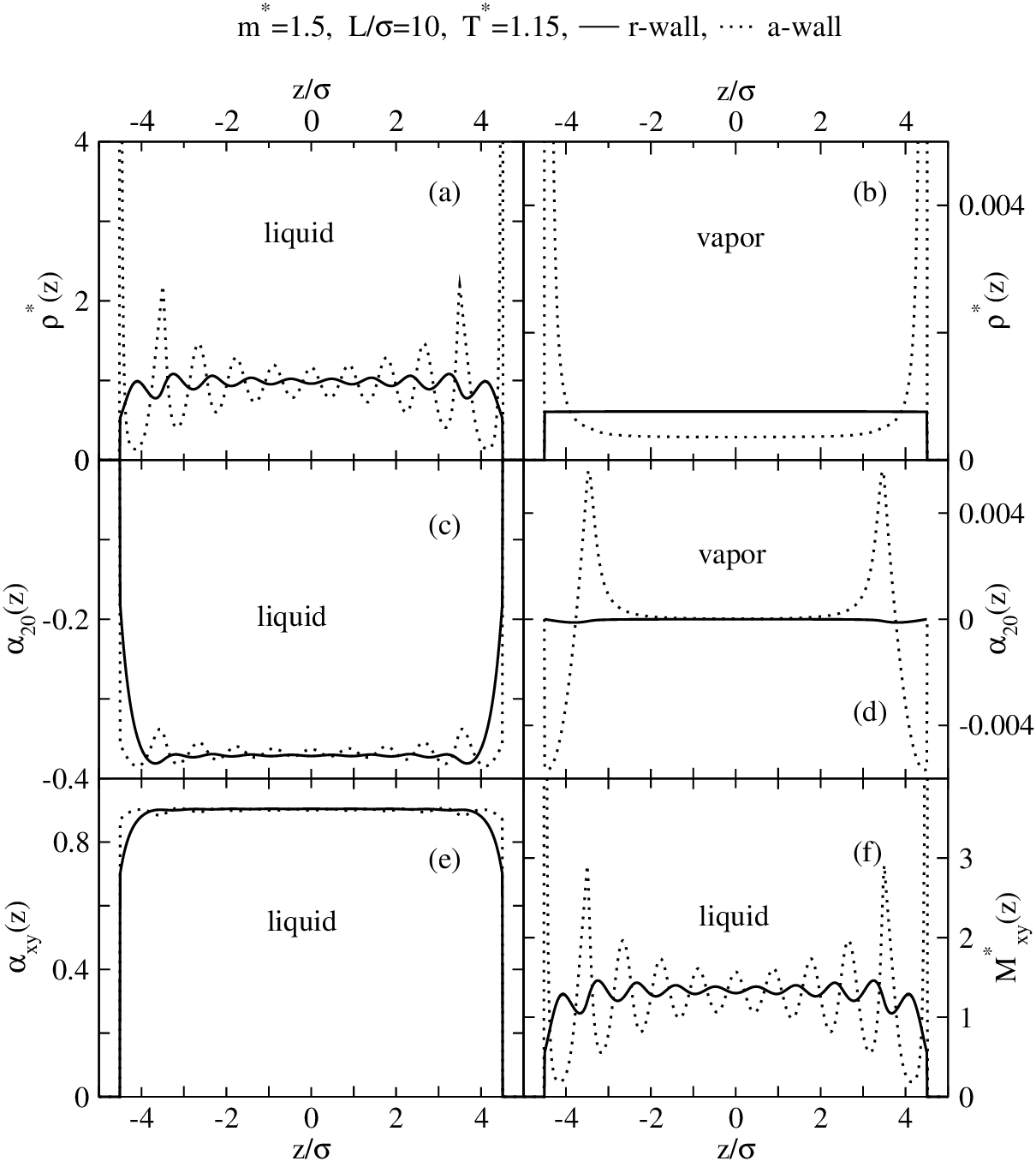}
\begin{spacing}{1.5}
\caption{\footnotesize Profiles of the number density $\rho^*(z)$ ((a) and (b)), preferential orientation $\alpha_{20}(z)$ of the dipole axes (Eq. (\ref{alma})) ((c) and (d)), ferromagnetic order parameter $\alpha_{xy}(z)$ (Eq. (\ref{axy})) (e), and in-plane magnetization $M^*_{xy}(z)=m^*\rho^*(z)\alpha_{xy}(z)$ (f) of coexisting ferromagnetic liquid and isotropic vapor phases for $m^*=1.5$, $L/\sigma=10$, and $T^*=1.15$, confined by {\it r}epulsive (---) or {\it a}ttractive (${\cdot}{\cdot}{\cdot}{\cdot}, \epsilon_w^*=1/2$) walls.}
\end{spacing}
\end{figure}
The density profiles of the coexisting isotropic gases 
are displayed in Fig. 5(b). For the attractive wall case the contact value of the density is $\rho^*_w\simeq0.55$ which indicates a visible spatial inhomogeneity in the vicinity of the walls. However, in the case of repulsive walls, at this low temperature the density of the vapor phase does not show any detectable inhomogenity. The behavior of the preferential orientation $\alpha_{20}(z)$ shown in Fig. 5(c) tells that in the liquid phase the dipole axes are preferentially oriented parallel to the walls throughout the pore for both types of confinements. The ferromagnetic ordering of the particles in the liquid phase is confirmed by the variation of the order parameter $\alpha_{xy}(z)$ (Eq. (\ref{axy})) which can be seen in Fig. 5(e) which describes the net orientation of the dipoles parallel to the walls, keeping in mind that for the orientational order parameter we find $\alpha_{10}(z)=0$ ($M_z=0$, see Eq. (\ref{us})) throughout of the pore. We note, that for all confined fluid phases studied here we found that $\alpha_{10}(z)=0$, i.e., throughout the pores the $z$ component of the magnetization $M_z$ is zero. Figure 5(f) displays the corresponding magnetization $M_{xy}(z)$ (Eq. (\ref{mxy})), which shows strong spatial inhomogeneities for both types of confinements induced by the structure of $\rho(z)$. (For the attractive walls the contact value of the magnetization is $M_{xy,w}\simeq25.7$.) For the vapor phase we find $\alpha_{xy}(z)=0$ for both attractive and repulsive confinements, which together with $M_z=0$ implies that the vapor phase is an isotropic phase. However, in Fig. 5(d) the order parameter $\alpha_{20}(z)$ shows that close to the walls there is a slight ordering in the vapor phases as well. There the ordering is more pronounced for the confinement by attractive walls than by repulsive walls. In the vapor phase, close to the attractive walls the dipole axes are preferentially ordered parallel to the walls ($\alpha_{20}<0$), but the amplitude of the orientation is much smaller than in the coexisting liquid phase (Fig. 5(d)). Between this layer of parallel orientation and the orientationally disordered interior of the pore (i.e., $\alpha_{20}(z)$ is close to zero), there is a layer with preferential orientation perpendicular to the wall ($\alpha_{20}>0$, Fig. 5(d)). With respect to the orientational order, our DFT results are consistent with those of Gramzow and Klapp \cite{kl1} but they show stronger spatial inhomogenities which is borne out by using a more sophisticated free energy functional for the hard sphere reference system.
\begin{figure}[t]
\includegraphics*[width=20pc]{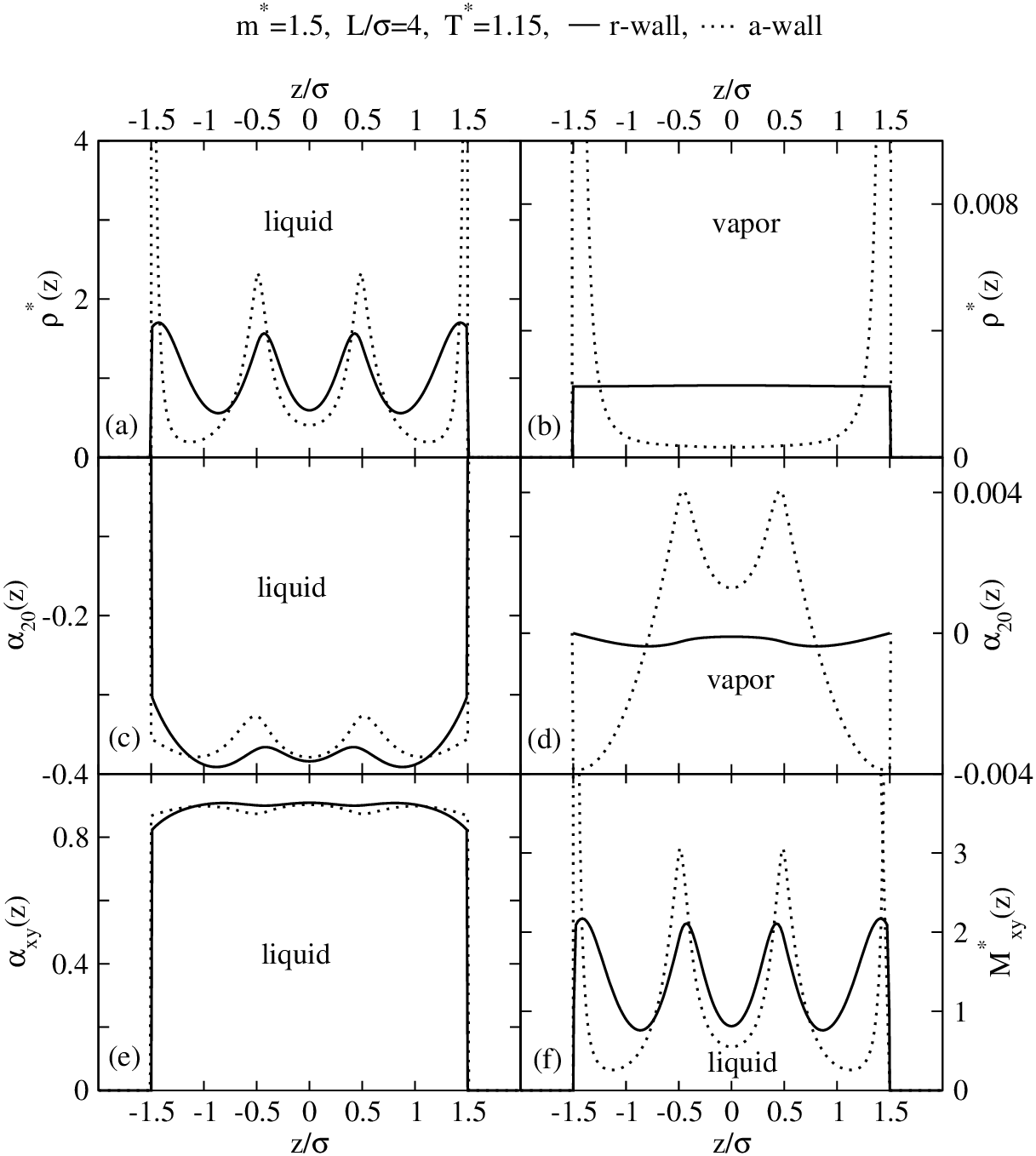}
\begin{spacing}{1.5}
\caption{\footnotesize Same as Fig. 5 for $L/\sigma=4$.}
\end{spacing}
\end{figure}
For $m^*=1.5$ and $T^*=1.15$, upon decreasing the wall separation from $L/\sigma=10$ 
to $L/\sigma=4$ the density and orientational distributions change significantly. 
Figure 6(a) shows that for both attractive and repulsive confinements the ferromagnetic liquid phases become more structured. Moreover, the confinement by attractive walls induces more ordered structures than by the corresponding repulsive ones. The density profiles for the coexisting isotropic vapor phase are displayed in Fig. 6(b), telling that only the attractive walls induce a spatial inhomogeneity in the density distribution of the isotropic vapor. 
\begin{figure}[t]
\includegraphics*[width=20pc]{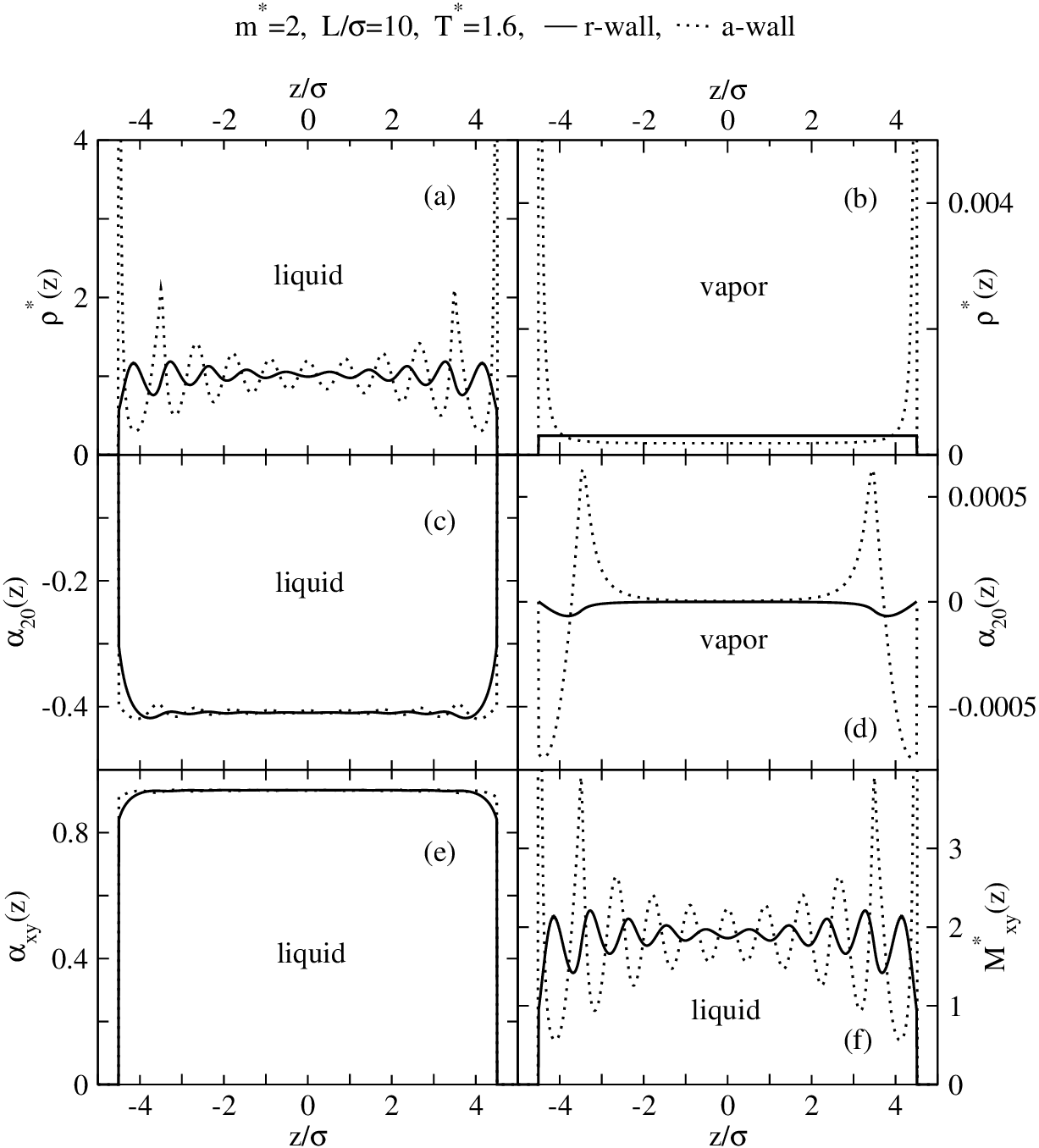}
\begin{spacing}{1.5}
\caption{\footnotesize Same as Fig. 5 for increased temperature ($T^*=1.6$) and stronger dipole moment ($m^*=2$).}
\end{spacing}
\end{figure}
\begin{figure}[t]
\includegraphics*[width=20pc]{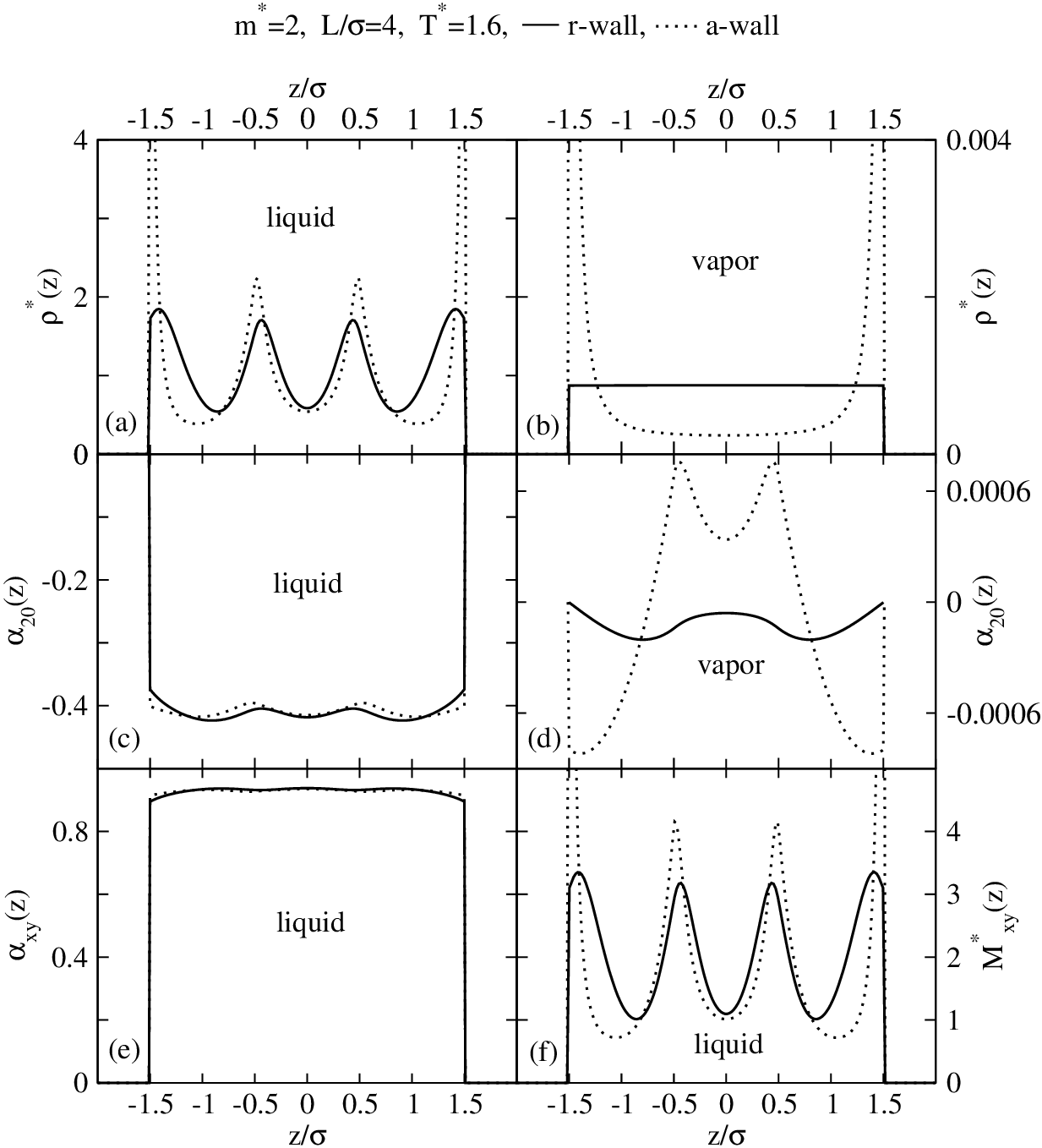}
\begin{spacing}{1.5}
\caption{\footnotesize Same as Fig. 7 for $L/\sigma=4$.}
\end{spacing}
\end{figure}
Figure 6(c) shows that in the liquid phase the extent of orientation of the dipole axes parallel to the walls is comparable with that in the wider ($L/\sigma=10$) pore but it varies more strongly across the pore. According to Fig. 6(d) the same conclusion holds for the structural properties of the confined vapor phase; since the vapor phase is isotropic this shows that the preferential ordering parallel and perpendicular to the walls does not hinge on the formation of ferromagnetic order.
From Fig. 6(e) one infers that the in-plane ferromagnetic order parameter of the liquid phase for $L/\sigma=4$ is comparable with that for $L/\sigma=10$. However, the thinner pore exhibits a stronger magnetization and more pronounced spatial variations (Fig. 6(f)).

Figure 7 displays the structural properties for the wide pore $L/\sigma=10$ at the higher temperature $T^*=1.6$ and for an increased dipole moment $m^*=2$. Compared with Fig. 5 ($T^*=1.15$, $m^*=1.5$) there are no qualitative differences. The quantitative differences concern the vapor density, which is lower due to the higher temperature (Fig. 7(b)), and the magnetization of the liquid phase, which is higher due to the increased dipole strength (Fig. 7(f)). The occurrence of preferential ordering parallel to the walls in the vapor phase ($\alpha_{20}(z)<0$, Fig. 7(d)) is in line with the findings of Gramzow and Klapp \cite{kl1}; however their approach does not render the preferential ordering perpendicular to the walls ($\alpha_{20}(z)>0$) in the subsequent layer towards the interior of the pore. But this preferential orientation is about a factor of ten smaller (Fig. 7(d)) as compared with the case shown in Fig. 5(d). 

For $m^*=2$ and $T^*=1.6$, decreasing the wall separation from $L/\sigma=10$ to $L/\sigma=4$ leads to a significant increase of both the liquid and the vapor densities (Figs. 8(a) and (b)) as well as to more pronounced density oscillation. The preferential orientional ordering is similar for the narrow (Figs. 7(c)-(e)) and the wide (Figs. 8(c)-(e)) pore. The magnetization is stronger in the more confined pore (Fig. 8(f)); this and the more pronounced spatial variation is caused by the corresponding behavior of the underlying number density (Fig. 8(a)). The vapor phase does not acquire a ferromagnetic order even in the narrow pore, i.e., $\alpha_{xy}(z)=0$ for the vapor phase.

\section{Summary}
In the present study of the phase behavior and structural ordering of confined Stockmayer fluids (Eq. (\ref{stm})) the 
following main results have been obtained.\\
1. We have applied an extension of the modified mean-field density functional theory to determine the phase behavior and the structure of Stockmayer fluids in slitlike pores formed by either attractive (Eq. (\ref{usw})) or repulsive (Eq. (\ref{uhw})) walls at wall separations $L/\sigma=10$ and $L/\sigma=4$, for reduced dipole moments $m^*=1.5$ and $m^*=2$, and for a reduced wall strength $\epsilon_w^*=1/2$. This system exhibits three distinct fluid phases: isotropic vapor, isotropic liquid, and ferromagnetic fluid. \\
2. For the reduced dipole moment $m^*=1.5$ we have found isotropic liquid {\textendash} isotropic vapor, ferromagnetic fluid {\textendash} isotropic vapor, and ferromagnetic fluid {\textendash} isotropic liquid first-order phase separations. The confinement by repulsive walls shifts the phase boundaries in the chemical potential {\textendash} temperature plane towards higher chemical potentials while attractive walls lead to a shift towards lower chemical potentials relative to the corresponding bulk data. The decrease of the wall separation from $L/\sigma=10$ to $L/\sigma=4$ does not change the topology of the phase diagrams, but in this case the shift of the phase boundaries relative to the bulk ones are more pronounced (see Figs. 1 and 2).
The increase of the dipole moment from $m^*=1.5$ to $m^*=2$ erases the isotropic liquid {\textendash} isotropic vapor phase transitions for both the confined and the bulk systems. We have found that for this stronger dipole moment only the phase boundary of the confined system with repulsive 
walls is significantly shifted relative to the bulk one (see Figs. 2 and 3). 
The phase boundaries of the bulk and of the confined system with attractive walls do intersect without a significant shift of the phase boundaries of the confined system relative to the bulk one.\\
3. We have shown that in the temperature {\textendash} chemical potential plane the Kelvin equation provides a good estimate of the confinement induced shift of the chemical potential for isotropic vapor {\textendash} ferromagnetic fluid (Fig. 1(a)) and isotropic fluid {\textendash} ferromagnetic fluid (Fig. 3(a)) phase coexistence. The multiple crossings of the isotropic fluid {\textendash} ferromagnetic fluid phase boundaries in the temperature {\textendash} chemical potential plane correspond to a non-monotonic temperature dependence of the contact angle of a drop of ferromagnetic fluid on a single wall. \\
4. On the basis of our DFT approach we have found that the number density $\rho(z)$ of the confined liquidlike phases (coexisting with the corresponding vaporlike phases) for both dipole moments ($m^*=1.5$ and $m^*=2$) and for both confinements ($L/\sigma=10$ and $L/\sigma=4$), with either attractive or repulsive walls, are strongly structured (see Figs. 5(a)-8(a)). The figures for the order parameter $\alpha_{20}(z)$ (Eq. (\ref{alma})) for preferential orientation of the dipole axes 
show that in the liquidlike phases the dipole axes are preferentially oriented parallel to the walls throughout the pore for both attractive and repulsive walls as well as for both sizes of confinements (see Figs. 5(c)-8(c)). The ferromagnetic ordering of the dipoles in the confined liquidlike phases is displayed by the variation of the ferromagnetic order parameter $\alpha_{xy}$ (see Eq. (\ref{axy}) and Figs. 5(e)-8(e)) which, together with the result $M_z(z)\equiv{0}$, describes the net orientation of the dipoles parallel to the walls. The corresponding magnetizations $M_{xy}(z)=m\rho(z)\alpha_{xy}(z)$ are displayed in Figs. 5(f)-8(f). Since $\alpha_{xy}(z)=0$ and $\alpha_{10}(z)=0$ throughout the pore for the vaporlike phases (at coexistence with the ferromagnetic liquidlike phases), they are completely isotropic. Attractive wall confinements give rise to strong adsorption at the walls even in the vaporlike phases (see Figs. 5(b)-8(b)). In the vaporlike phases, close to the attractive walls a slight preferential parallel ordering has been found. Between the layer of this preferential parallel orientation and the orientationally disordered interior of the pore a layer with preferential perpendicular orientation has been detected (see Figs. 5(d)-8(d)).
\begin{acknowledgments} 
I. Szalai acknowledges useful discussions with R. Roth and L. Harnau.
Financial support from the Hungarian Scientific Research Fund 
(Grant No. OTKA K61314) is also acknowledged.
\end{acknowledgments}
\appendix
\setcounter{section}{1}
\section*{Appendix A: Fundamental-measure theory}
One of the main limitations of the original version of fundamental-measure 
theory is that the underlying bulk fluid
equation of state reduces to the Percus-Yevick compressibility equation. 
As it is well known, 
within this approximation the contact value of the density profile of a 
one-component hard-sphere fluid at a planar wall is significantly 
overestimated at high bulk densities. In order to improve the 
quantitative accuracy and to maintain internal consistency, following 
Roth et al. \cite{ro3} and Tarazona \cite{ta1} here we apply a 
fundamental-measure theory based on the Carnahan-Starling equation of state. 
Within this approach the hard-sphere excess free energy density functional 
is given in terms of weighted densities as
\begin{equation}
f_{hs}=\frac{1}{\beta}{\Phi(\{n_{\alpha}(z)\})},
\label{fun1}
\end{equation}   
where
\begin{eqnarray}
\Phi=-n_0\ln(1-n_3)+\frac{n_1n_2-\mathbf{n}_1{\cdot}\mathbf{n}_2}{1-n_3}\nonumber\\
+(n_2^3-3n_2\mathbf{n}_2{\cdot}\mathbf{n}_2)\frac{n_3+(1-n_3)^2\ln(1-n_3)}
{36{\pi}n_3^2(1-n_3)^2}.
\label{bear}
\end{eqnarray}
In Ref. \cite{ro4} it is shown that the weighted densities in the present 
slab geometry are given as 
\begin{equation}
n_{\alpha}(z)=\int_{-\sigma/2}^{\sigma/2}{dz'\hat\rho(z+z')w^{(\alpha)}(z')},
\label{fun2}
\end{equation}
where the reduced weight functions $w^{(i)}$ as functions of $z$ are
\begin{eqnarray}
w^{(0)}=\frac{1}{\sigma},\;\;\;w^{(1)}=\frac{1}{2},
\;\;\;w^{(2)}=\pi\sigma,\;\;\;\nonumber\\
w^{(3)}={\pi}((\sigma/2)^2-z^2),\nonumber\\
\mathbf{w}^{(1)}=\frac{z\pi}{\sigma}\mathbf{e}_z,\;\;\;
\mathbf{w}^{(2)}=2{\pi}z\mathbf{e}_z.
\label{fun3}
\end{eqnarray}
The chemical potential functional (first-order direct correlation function) of the reference hard sphere system is 
\begin{eqnarray}
\beta\mu_{hs}[\{\hat\rho(z)\}]=\beta\frac{\delta{F_{hs}}[\{\hat\rho(z)\}]}{\delta\hat\rho(z)}\nonumber\\
=\sum_{\alpha=0}^5\int_{-L/2}^{L/2}d{z'}
\frac{\partial\Phi[\{\hat\rho(z')\}]}{\partial{n_{\alpha}}}
w^{(\alpha)}(z'-z),
\label{muf}
\end{eqnarray}
where the corresponding derivatives are
\begin{eqnarray}
\Phi_0=\frac{\partial\Phi}{\partial{n_0}}=-\ln(1-n_3),\nonumber\\
\Phi_1=\frac{\partial\Phi}{\partial{n_1}}=\frac{n_2}{1-n_3},\nonumber\\
\Phi_2=\frac{\partial\Phi}{\partial{n_2}}=\frac{n_1}{1-n_3}\nonumber\\
+(n_2^2-\mathbf{n}_2
\mathbf{n}_2)\frac{(n_3+(1-n_3)^2\ln(1-n_3))}{12\pi{n_3^2(1-n_3)^2}},\nonumber\\
\Phi_3=\frac{\partial\Phi}{\partial{n_3}}=\frac{n_0}{1-n_3}\nonumber\\
+\frac{n_1n_2-\mathbf{n}_1\mathbf{n}_2}{(1-n_3)^2}+\nonumber\\
(n_2^3-3n_2\mathbf{n}_2\mathbf{n}_2)\left(\frac{2+n_3(n_3-5)}{36\pi{n_3^2}(1-n_3)^3}
+\frac{\ln(1-n_3)}{18\pi{n_3^3}}\right),\nonumber\\
\Phi_4=\frac{\partial\Phi}{\partial{\mathbf{n}_1}}=-\frac{\mathbf{n}_2}{1-n_3},\nonumber\\
\Phi_5=\frac{\partial\Phi}{\partial{\mathbf{n}_2}}
=-\frac{\mathbf{n}_1}{1-n_3}\nonumber\\
-n_2\mathbf{n}_2\frac{n_3+(1-n_3)^2\ln(1-n_3)}{6\pi{n_3^2}(1-n_3)^2}.\;\;\;\;\;\;
\end{eqnarray}
In the summation (Eq. (\ref{muf})) $\alpha=4,5$ referes to the vectorial weight functions 
$\mathbf{w}^{(1)}$ and $\mathbf{w}^{(2)}$, respectively.

\appendix
\setcounter{section}{2}
\section*{Appendix B: Calculation of the coefficients $A^{(i)}_{l_1m_1l_2m_2}(x)$}
Solving the set of integral equations (\ref{rhoder}) and (\ref{alpder}) (see also Eqs. (\ref{rho1}) and (\ref{alp1})) requires to determine the functions $u^{(i)}_{l_1m_1l_2m_2}(z)$ (see Eq. (\ref{ukk})) which are based on the coefficients 
$A^{(i)}_{l_1m_1l_2m_2}(x)$ (see Eqs. (\ref{ukl2}) and (\ref{aikl})). By using Eq. (\ref{aikl}) we have calculated these coefficients and in the following we provide all coefficients which are nonzero for $i,l_1,l_2=0,1,2$:
\begin{eqnarray}
A^{(0)}_{0000}=8\pi^2,\;\;\;A^{(1)}_{1010}=\frac{8\pi^2}{3}(3x^2-1),\nonumber\\
A^{(1)}_{111\overline{1}}=-\frac{4\pi^2}{3}(3x^2-1),\nonumber\\
A^{(2)}_{0000}=\frac{16\pi^2}{3},\;\;A^{(2)}_{0020}=\frac{8\sqrt{5}\pi^2}{15}(3x^2-1),\nonumber\\
A^{(2)}_{2020}=\frac{8\pi^2}{5}(9x^4-8x^2+1),\nonumber\\
A^{(2)}_{212\overline{1}}=\frac{8\pi^2}{15}(-18x^4+15x^2-1),\nonumber\\
A^{(2)}_{222\overline{2}}=\frac{4\pi^2}{15}(9x^4-6x^2+1),
\end{eqnarray}
where
\begin{equation}
x{\equiv}\cos\theta_{12}=\frac{z}{\sqrt{R^2+z^2}}.
\end{equation}
The coefficients $A^{(i)}_{l_1m_1l_2m_2}$ are symmetric,
\begin{equation}
A^{(i)}_{l_1m_1l_2m_2}(x)=A^{(i)}_{l_2m_2l_1m_1}(x),
\end{equation}
so that the functions
\begin{equation}
u^{(i)}_{l_1m_1l_2m_2}(z)=u^{(i)}_{l_2m_2l_1m_1}(z)
\end{equation}
are symmetric, too, which simplifies Eqs. (\ref{rhoder}) and (\ref{alpder}).
We note that the integral in Eq. (\ref{ukl2}) splits into two parts 
due to the Heaviside function $\Theta{(R^2+z^2-\sigma^2)}$:
\begin{eqnarray}
\int_{0}^{\infty}{dR\,\Theta{(R^2+z^2-\sigma^2)}g(R,z)}\nonumber\\
=\Theta{(|z|-\sigma)}\int_0^{\infty}{dRg(R,z)}\nonumber\\
+\Theta{(\sigma-|z|)}\int_{\sqrt{\arrowvert{\sigma^2-z^2\arrowvert}}}^{\infty}dRg(R,z).
\end{eqnarray}
\appendix
\setcounter{section}{3}
\section*{Appendix C: Calculation of the density and orientational profiles}
Equations (\ref{rhoder}) and (\ref{alpder}) lead to coupled integral equations for 
$\rho(z)$ and $\alpha(z,\omega)$ (or $\alpha_{lm}(z)$, see Eq. (\ref{al}); here we suppress their dependences on $T$ and $\mu$):
\begin{widetext}
\begin{eqnarray}
\rho(z)=\exp\Big[-\beta\mu^{ref}_{hs}[{\{\rho(z)}\}]-\int{d\omega
{\alpha}(z,\omega)}\ln[4\pi\alpha(z,\omega)]\nonumber\\
\!\!\!\!\!\!\!\!\!\!\!\!\!\!\!\!\!\!\!\!\!\!\!\!\!\!\!\!\!\!\!\!
-2\beta\sum_{l_1,m_1}\alpha_{l_1m_1}(z)\sum_{l_2,m_2}
\int_{-L/2}^{L/2}{dz'}\rho(z')u_{l_1m_1l_2m_2}(z-z')\alpha_{l_2m_2}(z')+\beta\mu-\beta{u_w(z)}\Big].
\label{rho1}
\end{eqnarray}
Elementary calculations lead from Eq. (\ref{alpder}) to the following equation:
\begin{eqnarray}
\alpha(z,\omega)=\frac{\exp{[-2\beta\int_{-L/2}^{L/2}
{dz'\rho(z')}\sum_{l_1,m_1}
\alpha_{l_1m_1}(z')\sum_{l_2,m_2}u_{l_1m_1l_2m_2}(z-z')Y_{l_2m_2}^*(\omega)]}}
{\int{d\omega}\exp{[-2\beta\int_{-L/2}^{L/2}
{dz'\rho(z')}\sum_{l_1,m_1}
\alpha_{l_1m_1}(z')\sum_{l_2,m_2}u_{l_1m_1l_2m_2}(z-z')Y_{l_2m_2}^*(\omega)]}}
\label{alp1}
\end{eqnarray}
which fulfills the normalization requirement in Eq. (\ref{norm}). Using the expansion of $\alpha(z,\omega)$ in terms of spherical harmonics (Eq. (\ref{al})) one obtains 
for the coefficients $\alpha_{lm}(z)$ (Eq. (\ref{alma})):
\begin{eqnarray}
\alpha_{l_1m_1}(z)=
\frac{\int{d\omega}Y^*_{l_1m_1}(\omega)
\exp{[-2\beta\int_{-L/2}^{L/2}{dz'\rho(z')}\sum_{l_2,m_2}
\alpha_{l_2m_2}(z')\sum_{l_3,m_3}u_{l_2m_2l_3m_3}(z-z')Y^*_{l_3m_3}(\omega)]}}
{\int{d\omega}\exp{[-2\beta\int_{-L/2}^{L/2}
{dz'\rho(z')}
\sum_{l_2,m_2}
\alpha_{l_2m_2}(z')\sum_{l_3,m_3}u_{l_2m_2l_3m_3}(z-z')Y^*_{l_3m_3}(\omega)]}}.
\label{alp2}
\end{eqnarray}
\end{widetext}

Equations (\ref{rho1}) and (\ref{alp2}) are a set of coupled integral 
equations which are solved iteratively for the density profile $\rho(z)$ 
and the coefficients $\alpha_{lm}(z)$ of the orientational distribution.  
To this end, in Eq. (\ref{rho1}) the orientational entropy term must be expressed in terms of 
the coefficients $\alpha_{lm}(z)$. Using Eq. (\ref{alp1}), after some calculations 
one arrives at
\begin{widetext}
\begin{eqnarray}
\int{d\omega\alpha}(z,\omega)\ln[{4\pi\alpha}(z,\omega)]=\nonumber\\
\!\!\!\!\!\!\!\!\!\!\!\!\!\!\!\!\!\!\!
\ln[4\pi/C(z)]-2\beta\sum_{l_1,m_1}\sum_{l_2,m_2}\alpha_{l_1m_1}(z)
\int_{-L/2}^{L/2}dz\,\rho(z')u_{l_1m_1l_2m_2}(z-z')\alpha_{l_2m_2}(z'),
\end{eqnarray}
where
\begin{eqnarray}
C(z)=
{\int{d\omega}\exp{\left[-2\beta\int_{-L/2}^{L/2}{dz'\rho(z')}\sum_{l_1,m_1}
\alpha_{l_1,m_1}(z')\sum_{l_2,m_2}u_{l_1m_1l_2m_2}(z-z')Y_{l_2m_2}^*(\omega)\right]}}.
\end{eqnarray}
\end{widetext}
\setcounter{section}{0}



\section*{References}
\begin{thebibliography}{9}
\bibitem{ch1} Chandra N P, Ghosh S K 1997 {\it J. Chem. Phys.} {\bf 106} 2752
\bibitem{se1} Senapati S, Chandra A 2000 {\it J. Chem. Phys.} {\bf 112} 10467
\bibitem{kl1} Gramzow M, Klapp S H L 2006 {\it Phys. Rev.} E {\bf 75} 011605
\bibitem{we1} Richardi J, Pileni M P, Weis J J 2008 {\it Phys. Rev.} E {\bf 77} 061510
\bibitem{kl2} Jordanovic J, Klapp S H L 2008 {\it Phys. Rev. Lett.} {\bf 101} 038302
\bibitem{kl3} Trasca R A, Klapp S H L 2008 {\it Comp. Phys. Comm.} {\bf 179} 66
\bibitem{kl5} Trasca R A, Klapp S H L 2008 {\it J. Chem. Phys.} {\bf 129} 084702
\bibitem{ha1} Han K K, Cushman J H, Diestler 1993 {\it Mol. Phys.} {\bf 79} 537
\bibitem{bo1} Boda D, Chan K-Y, Henderson D 1998 {\it J. Chem. Phys.} {\bf 109} 7362
\bibitem{ex1} Matuo C Y, Bourdon A, Bee A, Figueiredo Neto A M 1997 {\it Phys. Rev.} E 
{\bf 56} R1310`
\bibitem{ex2} Vorobiev A, Major J, Dosch H, Gordeev G, Orlova D 2004 {\it Phys. Rev. Lett.} 
{\bf 93} 267203 
\bibitem{gr1} Groh B, Dietrich S 1994 {\it Phys. Rev.} E {\bf 50} 3814
\bibitem{wi1} Zhang H, Widom M 1994 {\it Phys. Rev.} E {\bf 49} 3591
\bibitem{gr2} Groh B, Dietrich S 2001 {\it Phys. Rev.} E {\bf 63} 021203
\bibitem{kl4} Range G M, Klapp S H L 2004 {\it Phys. Rev.} E {\bf 70} 031201
\bibitem{sz1} Szalai I and Dietrich S 2005 {\it Mol. Phys.} {\bf 103} 2873
\bibitem{ro3} Roth R, Evans R, Lang A, Kahl G 2002 {\it J. Phys.: Condens. Matter} {\bf 14} 12063
\bibitem{ev1} Evans R 1979 {\it Adv. Phys.} {\bf 28} 143
\bibitem{ro1} Rosenfeld Y 1989 {\it Phys. Rev. Lett.} {\bf 63} 980
\bibitem{ro2} Rosenfeld Y 1990 {\it Phys. Rev.} A {\bf 42} 5978
\bibitem{te1} Teixeira P I, Telo da Gama M M 1991 {\it J. Phys.: Condens. Matter} {\bf 3} 111
\bibitem{fr1} Frodl P, Dietrich S 1992 {\it Phys. Rev.} A {\bf 45} 7330
\bibitem{fr2} Frodl P, Dietrich S 1993 {\it Phys. Rev.} E {\bf 48} 3741
\bibitem{di1} Dietrich S 1988 in {\it Phase Transitions and Critical Phenomena} 
(Domb C and Lebowitz J L, eds.) Vol. {\bf 12}, 1-218 (Academic Press, London)
\bibitem{ta1} Tarazona P 2002 {\it Physica} A {\bf 306} 243
\bibitem{ro4} Roth R, Dietrich S 2000 {\it Phys. Rev.} E {\bf 62} 6926
\end{thebibliography}
\end{document}